%% file: syncprims.tex
\documentclass{acm_proc_article-sp}

\def\unnumfootnote{\xdef\@thefnmark{}\@footnotetext}

\usepackage{algorithm}
\usepackage{algorithmic}
\usepackage{multicol}
\usepackage[scaled=.85]{helvet}
\usepackage{mathptmx}
\usepackage{cite, wrapfig, multirow, url}
\usepackage{graphicx}
\usepackage{parskip}
\usepackage{amsmath}
\usepackage{verbatim}
\usepackage{booktabs}
\usepackage{url}
\usepackage{color}
\usepackage[small]{caption}
\DeclareCaptionType{copyrightbox}
\newcommand{\code}[1]{\emph{#1}}

\providecommand{\shortcite}[1]{\cite{#1}}
\begin{document}
%
% --- Author Metadata here ---
\conferenceinfo{arxiv.org}{2011 October 20}
%\CopyrightYear{2007} % Allows default copyright year (20XX) to be over-ridden - IF NEED BE.
%\crdata{0-12345-67-8/90/01}  % Allows default copyright data (0-89791-88-6/97/05) to be over-ridden - IF NEED BE.
% --- End of Author Metadata ---

\title{Efficient Synchronization Primitives for GPUs}

% \numberofauthors{1}
\numberofauthors{1}

\author{
% \alignauthor
% Double-Blind Review Authors
% \alignauthor
% % 1st. author
\alignauthor
Jeff A. Stuart\\
       \affaddr{Department of Computer Science}\\
       \affaddr{UC Davis}\\
       \email{stuart@cs.ucdavis.edu}
\and
% 2nd. author
\alignauthor
John D. Owens\\
       \affaddr{Department of Electrical and Computer Engineering}\\
       \affaddr{UC Davis}\\
       \email{jowens@ece.ucdavis.edu}
}

\maketitle

\begin{abstract} \input{abstract} \end{abstract}



\section{Introduction}            \label{sec:intro}             \input{intro}
\section{Background}              \label{sec:background}        \input{background}
\section{Benchmarks}              \label{sec:benchmarks}        \input{benchmarks}
\section{Machine Abstraction}     \label{sec:machine_model}     \input{machine_model}
\section{Design}                  \label{sec:design}            \input{design}
\section{Results \& Discussion}   \label{sec:results}           \input{results}
\section{Conclusion}              \label{sec:conclusion}        \input{conclusion}
\bibliographystyle{abbrv}
\bibliography{bib/all}
\appendix                           \label{sec:appendix}           \input{appendix}

\end{document}

%% file: abstract.tex
In this paper, we revisit the design of synchronization primitives---specifically barriers, mutexes, and semaphores---and how they apply to the GPU\@. Previous implementations are insufficient due to the discrepancies in hardware and programming model of the GPU and CPU\@. We create new implementations in CUDA and analyze the performance of spinning on the GPU, as well as a method of sleeping on the GPU, by running a set of memory-system benchmarks on two of the most common GPUs in use, the Tesla- and Fermi-class GPUs from NVIDIA\@. From our results we define higher-level principles that are valid for generic many-core processors, the most important of which is to limit the number of atomic accesses required for a synchronization operation because atomic accesses are slower than regular memory accesses. We use the results of the benchmarks to critique existing synchronization algorithms and guide our new implementations, and then define an abstraction of GPUs to classify any GPU based on the behavior of the memory system. We use this abstraction to create suitable implementations of the primitives specifically targeting the GPU\@, and analyze the performance of these algorithms on Tesla and Fermi. We then predict performance on future GPUs based on characteristics of the abstraction. We also examine the roles of spin waiting and sleep waiting in each primitive and how their performance varies based on the machine abstraction, then give a set of guidelines for when each strategy is useful based on the characteristics of the GPU and expected contention.

%% file: intro.tex
The general-purpose application space of the GPU is rapidly expanding beyond the kernels of data-parallel scientific applications into workloads with substantially more control complexity. Demanding, control-complex tasks such as irregular memory access and control flow~\cite{Zhang:2011:OEO}, backtracking search~\cite{Jenkins:2011:LLF}, and task queuing~\cite{Kogan:2011:WQW, Tzeng:2010:TMF} are all pushing the GPU into new application domains.

\let\thefootnote\relax\footnotetext{\input{reader_note}}

In the CPU world, to address application areas such as these in a parallel way, programmers would commonly and frequently synchronize between parallel threads. Fortunately, the space of CPU synchronization primitives is well-studied (Section~\ref{sec:background}), and high-performance primitives such as mutexes, barriers, and semaphores are part of any modern operating system.

On the GPU, however, we focus our work on synchronizations between blocks of threads, and this is uncharted territory. Why? The first reason is that atomic operations, necessary on the GPU for most inter-block synchronizations, were not even part of the capabilities of the first general-purpose GPUs (NVIDIA's G80 family, and more generally NVIDIA's Compute Version 1.0 GPUs, support no atomic instructions). However, the recent addition of atomic operations, and the rapid improvement of their performance, in both NVIDIA's and AMD's GPUs provides the substrate for high-performance synchronization primitives.

The second reason is our focus in this paper: researchers have not yet considered how synchronization primitives should be designed and optimized for performance on the GPU\@. The architecture of the GPU differs significantly from the CPU, and so we expect, and show in this paper, that GPU synchronization primitives differ in significant ways from their CPU cousins. Today, by far the most common approach to synchronization on the GPU is the lowly spin-lock, which suffers from two problems: it is poorly suited for high-performance synchronization, and it is a low-level primitive whereas programmers would prefer a higher level of abstraction.

We take a systematic approach to the design and optimization of high-performance synchronization primitives, building a set of benchmarks to analyze the memory system and derive a performance model for a GPU memory system to make better design decisions when implementing these primitives. The main contributions of this research are the classification of GPUs based on a machine abstraction, the benchmarks used for the classification, a high-level library for synchronization primitives on the GPU, and making the same library high-performance by replacing as many atomics as possible with non-atomic operations (thus gaining measurable speedup), and a thorough analysis of the library on two different architectural families of NVIDIA GPUs.

%% file: reader_note.tex
If you are reading this paper, we hope that you have an interest in synchronization primitives on the GPU. Furthermore, we hope that you have applications that might benefit from such primitives. If so, please contact the authors. We are actively seeking real-world and research applications that have a need for such primitives, and would be happy to work with you on incorporating these primitives into your system.

%% file: background.tex
Synchronization primitives such as barriers, mutexes, and semaphores (all briefly explained in Section~\ref{sec:design}) are essential to parallel computing. especially when a multi-core/many-core machine must provide coherent access to a shared resource such as a work or task queue.

\subsection{CPU Primitives}

Perhaps the most common form of synchronization in parallel programming today is the barrier. On shared-memory machines, centralized barriers are quite common as they have a simple implementation. However, they do not scale well with the number of threads, so researchers devised decentralized implementations such as butterfly barriers~\cite{Brooks:1968:TBB} and hierarchical (tree) barriers~\cite{Arenstorf:1989:CBA}.

Mutexes are another common primitive used for synchronization on machines with more than one processing core. Significant time and effort over the past several decades has been put into researching new locking mechanisms for mutexes. The most naive approach, given access to atomic instructions, is a simple spin lock, which is the method used exclusively in GPU research~\cite{Tzeng:2010:TMF, Aila:2009:UTE} and demonstrated in Algorithm~\ref{listing:spinlock}. Spin locks have several drawbacks; first and foremost is that they cannot guarantee fairness. Secondly, they scale poorly with the number of threads due to their impact on the memory system. To alleviate the problems of spin locks, researchers devised other methods. Anderson implemented backoff algorithms similar to the collision-avoidance capabilities in Ethernet~\cite{Metcalfe:1976:EDP} for spin locks to help reduce contention~\cite{Anderson:1990:TPO}. Anderson et al.\ identified the performance implications with waiting and blocking (also called sleeping) on SMP architectures~\cite{Anderson:1989:TPI}. Ousterhout proposed scheduling techniques such as busy-waiting for a time and then sleeping~\cite{Ousterhout:1982:STF}. Boguslavsky et al.\ then devised a model to give optimal strategies for spinning and blocking~\cite{Boguslavsky:1993:OSF}. Still, these methods did not guarantee fairness. Mellor-Crummey and Scott devised more mutex algorithms that both lowered contention of the memory system and offered starvation prevention~\cite{Mellor:AFS:1991}. Such algorithms include the ``ticket'' system, also known as ``fetch-and-add'', and array- and list-based queuing. The modern Linux kernel uses a more advanced implementation of a mutex called a Futex (short for ``Fast User-space Mutex''), which is a combination of spin-waiting in user space followed by the kernel putting threads to sleep when the lock is under contention~\cite{Russell:2002:FFF}.

Semaphores are a lesser-used primitive, but still important in certain applications. A semaphore guards access to a critical section. It is similar to a mutex, but it allows a user-specified maximum number of threads in to a critical section simultaneously. The two operations of a semaphore are \code{wait()} (causes a thread to block while the semaphore is at capacity) and \code{post()} (reduces the current logical count of threads within the critical section by one, thereby granting access to a single waiting thread). Dijkstra was the first to publish research on semaphores~\cite{Dijkstra:1968:CSP}.

% *** Something more recent would be highly desirable here for final submission

\subsection{GPU Primitives}

The GPU is a many-core machine. Threads are grouped into a set of blocks, and each block is scheduled onto one of the streaming multiprocessors (SM) and run until completion. Many of the CPU implementations for barrier do not directly port to the GPU (they use instructions not available on the GPU), or do not scale well on the GPU because they are meant for different types of processors and different memory systems than that of the GPU\@. NVIDIA provides intra-block barriers in the form of highly efficient intrinsics (e.g. \code{\_\_syncthreads()}), but does not provide a software mechanism for inter-block barriers. The typical method for such a barrier is to simply end one kernel and begin another, triggering a global synchronization, but this is quite slow compared to other software methods\footnote{Volkov and Demmel, in their 2008 paper~\cite{Volkov:2008:BGT}, noted that a GPU synchronous kernel invocation took 10--14~$\mu$s and a GPU asynchronous invocation took 3--7~$\mu$s.}. Xiao and Feng recently showed that implementations of CPU-friendly barriers, such as hierarchical and decentralized barriers, are slower on the GPU than just a simple two-stage atomic barrier~\cite{Xiao:2010:IBG}. Their design of a scalable and decentralized global barrier (without atomics) outperforms all other known barrier implementations on the GPU\@ by between $3\times$ and $7\times$. Their success in designing and optimizing a fast GPU barrier inspires our work here.

Like barriers, CPU mutex algorithms tend to be ill-suited for the GPU\@. Many of the most common CPU implementations involve linked structures (e.g.\ a linked list) of some kind, and some sort of spin-then-block approach. On the GPU, these implementations pose two specific problems: the GPU is not well-equipped to handle linked structures (linked structures often cause divergence in warps and require small non-coalesced reads, both of which impact performance much more on a GPU than on a CPU) and there is no blocking mechanism on the GPU\@. As such, any mention of mutual exclusion in GPU programming tends to point to aggressive spin locks using atomics or to poor-performance software workarounds~\cite{Shams:2007:EHA} that require a large (or potentially even unbounded) number of global-memory writes to ensure correctness. Beyond the fact that programmers use an inefficient method for spin locking, the fact that they simply use spin locks and not a higher-level abstraction such as a mutex shows the immaturity of synchronization primitives on the GPU\@. Unlike barriers and mutexes, the authors are unaware of any research regarding semaphores on the GPU\@.

\newcommand{\proc}{\textbf{procedure}\ }
\newcommand{\func}{\textbf{function}\ }
\newcommand{\ATRUE}{\algorithmictrue{}}
\newcommand{\AFALSE}{\algorithmicfalse{}}

\begin{algorithm}

  \func CPU: CreateSpinLock
    \begin{algorithmic}[1]
      \STATE   \textit{X} $\leftarrow$ AllocateGPUWord()
      \STATE   *\textit{X} $\leftarrow 0$
      \RETURN  \textit{X}
    \end{algorithmic}

  \vspace{5pt}

  \func GPU: SpinLock(Lock)
    \begin{algorithmic}[1]
      \STATE \textit{Locked} $\leftarrow \AFALSE$
      \WHILE{\textit{Locked} $= \AFALSE$}
        \STATE \textit{OldVal} $\leftarrow$ atomicExch(\textit{Lock}, 1)
        \IF{\textit{OldVal} $ = 0$}
          \STATE \textit{Locked} $\leftarrow \ATRUE$
        \ENDIF
      \ENDWHILE
    \end{algorithmic}

  \vspace{5pt}

  \func GPU: SpinUnlock(Lock)
    \begin{algorithmic}[1]
      \STATE: atomicExch(\textit{Lock}, 0)
    \end{algorithmic}

  \caption{A spin lock implemented on NVIDIA GPUs using CUDA\@. On the CPU, the user first allocates a word of memory and sets it to zero. On the GPU, to acquire the lock, a thread will simply continue to atomically exchange the value of the lock with 1. If the old value is ever 0, it means the lock is free and the thread then just acquired the lock. To return the lock, a thread simply sets the value back to 0. This is a less-than-ideal implementation, especially on the GPU, due to the high atomic contention involved in acquiring the lock. Also, we use \code{atomicExch()} instead of a volatile store and \code{threadfence()} because the atomic queue has predictable behavior, \code{threadfence()} does not (i.e.\ it can vary greatly in execution time if other memory operations are pending).\label{listing:spinlock}}

\end{algorithm}

%% file: benchmarks.tex
\begin{table*}
  \noindent
  \begin{center}
    \begin{small}
      \begin{tabular}[th]{lrrrr}
        \toprule
                                                    & Tesla Reads (ms) & Tesla Writes  (ms) & Fermi Reads (ms) & Fermi Writes (ms) \\
        \midrule
        Contentious Volatile                        &  0.848    &  0.829      & 0.494     & 0.175 \\
        Noncontentious Volatile                     &  0.590    &  0.226      & 0.043     & 0.029 \\
        Contentious Atomic                          & 78.407    & 78.404      & 1.479     & 1.470 \\
        Noncontentious Atomic                       &  0.845    &  0.991      & 0.437     & 0.312 \\
        Contentious Volatile preceded by Atomic     &  0.923    &  0.915      & 1.473     & 0.824 \\
        Noncontentious Volatile preceded by Atomic  &  0.601    &  0.228      & 0.125     & 0.050 \\
        \bottomrule
      \end{tabular}
    \end{small}
  \caption{Times for different numbers and types of memory accesses on Tesla and Fermi. The results for each were obtained by performing one thousand memory accesses per block with a fully saturated GPU (240 blocks on Tesla, 128 blocks on Fermi). The comparison of results on the same GPU is important, the comparison of absolute results between Tesla and Fermi does not matter. With contentious memory accesses, each block accesses the exact same four-byte word continuously. With noncontentious memory accesses, each block accesses its own unique four-byte word continuously. All accesses are cached on Fermi, and none are cached on Tesla (Tesla has no cache). \label{table:benchmark_results}}
   \end{center}
\end{table*}

We set out to design a set of primitives that would be useful for applications on the GPU that require inter-block synchronization. We realized that while doing this, we would need to investigate a variety of algorithms. As hardware is diverse across vendors and even across different GPUs from the same vendor, we expect differences in both absolute performance and performance trends across all GPUs. We wish to study this variation and pick two GPUs---the GT200 class of Tesla and the GF100 class of Fermi, currently the two most commonly used GPUs for computing.

To gain a better understanding of which algorithms to investigate, and to gain insight into possible new implementations, we wanted to devise a machine abstraction. We could then classify our GPUs based on this abstraction, paying close attention to the performance of the memory system, which we believe will be the most common bottleneck for synchronization primitives on the GPU\@.

To develop a model for each GPU, we implemented a set of twelve benchmarks to test the performance of memory systems. At a high level, the benchmarks can be classified into \emph{atomic} memory accesses and \emph{volatile} non-atomic memory accesses, and then further divided into \emph{contentious} accesses and \emph{noncontentious} accesses. At the final level, we divide the sets into \emph{read/load} benchmarks and \emph{write/store} benchmarks.

The GPU is made up of a number of shared multiprocessors (SMs, essentially vector processing units). Each SM can hold up to a fixed number of blocks at any one time (both GPUs we consider allow up to eight), depending on kernel resource usage (e.g. registers per block), so we can saturate the SMs and achieve maximum memory bandwidth with very few blocks. Thus we do not have to worry about block start-up/spin-down costs affecting our timings.

The GPU already has very fast intra-block synchronization intrinsics, and it makes little sense to have hundreds of threads in the same block contending for the same primitive. Hence the work we describe here has \emph{block semantics}, meaning an entire block (or only one thread within a block) acquires a mutex/semaphore rather than each individual thread (note this does not preclude a block from holding more than one resource at a time). Block semantics are common in GPU programming; because inter-block synchronization is much more expensive than intra-block synchronization, it is more efficient to have only one request per block then use the faster intra-block synchronization mechanisms to arbitrate between threads in a block.

Our benchmarks test the memory system when memory locations are accessed by only a single thread per block (we refer to the selected thread as the master thread). In each benchmark, the master thread simply performs its type of memory access (e.g.\ contentious atomic read) one thousand times. For contentious memory accesses, every master thread continuously accesses the same four-byte word. For noncontentious accesses, they repeatedly access their own unique four-byte word that lies within its own 256-byte boundary, so the words are not on the same memory line\footnote{We would like to thank NVIDIA architect \textit{anonymous} for his valuable insights into the GPU hardware, especially the memory system and atomic units.}. Volatile loads (VLs) and stores (VSs) are self-explanatory. For atomic reads, we use \code{atomicAdd(memory, 0)} and for atomic writes, we use \code{atomicExch(memory, 0)}.

We implemented an additional four tests beyond the aforementioned eight. Because atomics are slower than volatile memory accesses, it could be beneficial to limit the number of atomic accesses necessary and perform all/as much of the necessary atomic accesses as early as possible, and to then rely only on volatile accesses at the end of the algorithms. However, as atomics essentially serialize memory transactions, we wanted to investigate what happens to the memory system when a volatile access immediately follows an atomic access. To do so, we replicate our four volatile memory tests, but execute one single atomic instruction at the start of execution for each block.

The absolute results from our benchmarks are not the most important results. Instead, it is the ratios between contentious and noncontentious accesses, as well as the ratios between atomic and volatile accesses, that really matter. These ratios will dictate which class of algorithms is best suited for each GPU\@.

\begin{table*}
  \noindent
  \begin{center}
    \begin{small}
      \begin{tabular}[th]{lrrrr}
        \toprule
                                       & Tesla Reads    & Tesla Writes  & Fermi Reads   & Fermi Writes \\
        \midrule
        Volatiles                      &  1.44$\times$  &  3.67$\times$ & 11.49$\times$ &  6.03$\times$ \\
        Atomics                        & 92.79$\times$  & 79.12$\times$ &  3.38$\times$ &  4.71$\times$ \\
        Volatiles preceded by Atomic   &  1.54$\times$  &  4.01$\times$ & 11.78$\times$ & 16.48$\times$ \\
        \bottomrule
      \end{tabular}
    \end{small}
  \caption{Ratios of time required for contentious accesses to noncontentious accesses. Fermi and Tesla results are compared to different baselines. Fermi results are 128k (one thousand per block on a saturated GPU) contentious accesses compared to the same number of noncontentious Fermi accesses, Tesla results are 240k (one thousand per block on a saturated GPU) contentious accesses compared to the same number of noncontentious Tesla accesses. All accesses are cached on Fermi, none are cached on Tesla because Tesla has no cache. Contentious Tesla atomics are very expensive compared to all other reads. Fermi atomics have much better behavior under contention, but are still slower than noncontentious accesses. On Fermi, but not on Tesla, the atomic unit seems to serialize all transactions on the same line until no more pending transactions exist. Under contention, this yields a cascading effect, turning all accesses into serialized (atomic) accesses. \label{table:benchmark_ratios_1}}
   \end{center}
\end{table*}

\begin{table*}
  \noindent
  \begin{center}
    \begin{small}
      \begin{tabular}[th]{lrrrr}
        \toprule
                                                    & Tesla Reads    & Tesla Writes  & Fermi Reads   & Fermi Writes \\
        \midrule
        Contentious Atomics                         & 92.46$\times$  & 94.57$\times$ &  2.99$\times$ &  8.40$\times$ \\
        Noncontentious Atomics                      &  1.43$\times$  &  4.38$\times$ & 10.16$\times$ & 10.76$\times$ \\
        Contentious Volatile preceded by Atomic     &  1.08$\times$  &  1.10$\times$ &  2.98$\times$ &  4.71$\times$ \\
        Noncontentious Volatile preceded by Atomic  &  1.02$\times$  &  1.01$\times$ &  2.91$\times$ &  1.72$\times$ \\
        \bottomrule
      \end{tabular}
    \end{small}
    \caption{Ratios of time required compared to volatile accesses. All Fermi results represent 128k (one thousand per block on a saturated GPU) memory accesses compared against the same number of volatile accesses (again, on Fermi). All Tesla results represent 240k (one thousand per block on a saturated GPU) memory accesses compared against the same number of volatile accesses (again, on Tesla). The most important trends in this information are 1) contentious atomics on Fermi are far less punishing than on Tesla, and 2) the atomic units on Tesla do not serialize volatile accesses immediately following atomic accesses. \label{table:benchmark_ratios_2}}
   \end{center}
\end{table*}

\paragraph*{\textbf{Benchmark Results}}

We present the full results of our benchmarks in Tables~\ref{table:benchmark_results}, \ref{table:benchmark_ratios_1}, and \ref{table:benchmark_ratios_2}.\ Tesla has no L2 cache (only a user-managed software cache that is private to each individual block), thus the atomic units retrieve operands from DRAM\@. On Fermi, the atomic units retrieve operands from the L2 cache (assuming the operand is in cache, otherwise it is moved from DRAM into L2). While it is obvious that atomic accesses are slower than volatile accesses, the magnitude by which they are slower differs dramatically between Tesla, with a worst case of approximately 90x slower, and Fermi, with a worst case of approximately 3x slower. And while we expect contentious accesses to be slower than noncontentious accesses, Fermi significantly speeds up noncontentious accesses when compared to Tesla. On Fermi, noncontentious volatile reads and writes are approximately 10x and 5x faster respectively than their contentious volatile counterparts. On Tesla, these ratios drop to approximately 1.3x and 3.5x respectively.

Another interesting difference between Tesla and Fermi has to do with the last four benchmarks we described above, where each master thread issues a single atomic access before its sequence of volatile accesses. On Tesla, these benchmarks had virtually no difference between the benchmarks that issued only volatile accesses. However, on Fermi, the performance degraded noticeably, and in some cases to the point where the total time was the same as that of the benchmarks with only atomic accesses. Volatile loads issued while an atomic unit still has control of the memory line are serialized by the atomic unit, essentially treating them as an \code{atomicAdd(memory, 0)}. We saw similar, but not quite as poor, results for writes.

% \begin{table*}
%   \noindent
%   \begin{center}
%     \begin{small}
%       \begin{tabular}[th]{lllll}
%         \toprule
%                                                     & Tesla vs Fermi \\
%         \midrule
%         Atomics                                     & Fermi significantly reduces the gap between volatile accesses and atomics compared to Tesla. \\
%         Contentious vs. Noncontentious Volatile     & The ratio with Tesla is better than with Fermi. \\
%         Atomic Unit Holds Lines Hostage?            & On Tesla, no. On Fermi, yes. \\
%         \bottomrule
%       \end{tabular}
%     \end{small}
%     \caption{Summary of memory-system behavior under different loads on both Tesla and Fermi, as examined with our machine abstraction in mind. \label{table:benchmark_ratios_3}}
%    \end{center}
% \end{table*}

%% file: machine_model.tex
Using the results from our benchmarks, we can now abstract the GPU, and specifically the memory system, with respect to the implementation of high-performance synchronization primitives. This abstraction gives us a guide to acceptable tradeoffs we can make in our algorithm designs, and describes a GPU in terms of its performance in executing synchronization primitives. Our abstraction narrows the  many parameters we could use to define the model to what we consider to be the three most important. All three parameters characterize the memory system, as synchronization primitives have virtually no compute requirements but make heavy use of the memory system. The characteristics we choose for our abstraction are:

\begin{description}

  \item[Atomic:Volatile]\hspace{-2pt}\textbf{memory access performance ratio.} This is the most important characteristic, especially under contention. It dictates whether simple implementations such as spin locks, which require atomics, are viable compared to other designs.

  \item[Contentious:Noncontentious]\hspace{-2pt}\textbf{volatile access ratio.} This determines how well a ``sleeping'' algorithm performs.

  \item[Do atomic units]\hspace{-2pt}\textbf{with a non-empty queue hold a line hostage?} This will dictate what types of ``sleeping'' implementations are viable (e.g.\ volatile polling on a memory location that is updated via atomics).

\end{description}

The rest of this section describes high-level strategies for implementing high-performance synchronization primitives on our two target GPUs, and the following section dives into the detail of our design choices and implementations for each of our target primitives.

\paragraph*{\textbf{Tesla}}

Tesla is characterized by very slow contentious atomics compared to both contentious and noncontentious volatile accesses. Reducing contention can theoretically yield almost two orders of magnitude in speedup. The speed discrepancies between contentious and noncontentious volatile accesses are insignificant in comparison. Preceding volatile accesses with an atomic (testing to see if an atomic unit holds lines hostage) does not noticeably degrade performance.

These facts imply that certain trade offs are acceptable and point to optimal implementation strategies. Any contentious atomics yield poor performance. Spin locks are clearly a poor decision, as are centralized atomic barriers. Implementations that can avoid atomic contention, and perhaps substitute it with VL/VS contention, seem ideal. Algorithms that spin block on a volatile read should yield great performance (of course, this also depends on the stress of the memory subsystem from other blocks), but those that use backoff with atomics will most likely be too unpredictable in their behavior because volatile accesses must go to DRAM, and the overhead varies, much more so than cached L2 accesses on Fermi, based on system load.

\paragraph*{\textbf{Fermi}}

The speed of the Fermi atomic unit is much better than that of Tesla, both in terms of raw time, and in terms of comparison to volatile memory accesses. However, Fermi has the disadvantage that an atomic unit holds its line and serializes all accesses until it successfully flushes its entire queue. Of course on any GPU, contentious volatile accesses are slower than noncontentious. However, the performance ratio of contentious to noncontentious volatile accesses is worse on Fermi than Tesla. This seems to be primarily due to the cache on Fermi, where serialization of requests for lines will have a more noticeable impact. Coherence could also impact this, as maintaining coherence across the cache requires flushes.

As contentious atomics are not even an order of magnitude slower than contentious volatile accesses, spin locks are a tempting solution. With the speed of the Fermi cache, a backoff algorithm should yield speedup as it reduces/eliminates contention without excessive waiting times. Performing atomics up front and then switching to volatile accesses will not yield the same performance benefit on Fermi because the atomic units can hold cache lines hostage and serialize requests.

%% file: design.tex
Using the results from our GPU abstraction, we want to explore existing algorithms and new algorithms for synchronization primitives. We limit the scope to algorithms that have a good chance of working well on the GPU\@. This section describes our design alternatives; the next section explores their performance. In order to frame the discussion, we define a few terms.

\begin{description}

  \item[Barrier] A synchronization primitive that guarantees all participating threads/blocks reach a specific point in code before any thread/block may progress beyond that point.

  \item[Mutex] A synchronization primitive that guarantees mutual exclusion and serialized access to a critical section, which is accomplished via \code{lock()} and \code{unlock()} methods.

\item[Semaphore] A synchronization primitive that guarantees that no more than $n$ threads/blocks can access a critical section simultaneously, which is accomplished via \code{wait()} (similar to \code{lock()}) and \code{post()} (similar to \code{unlock()}) methods.

  \item[Spinning (CPU)] When a thread continuously monitors for a change of state by polling a memory location.

  \item[Spinning (GPU)] When a thread uses processor time to simultaneously wait for a change of state and then modify that state. This is done by aggressively accessing a memory location using an atomic operation (e.g.\ \code{atomicExch()}).

  \item[Sleeping/Blocking (CPU)] When the OS puts a thread to sleep until a certain condition has been met. This frees up all processor-specific resources consumed by a thread and prevents the thread from receiving processor time.

  \item[Sleeping/Blocking (GPU)] When a thread or block polls a memory location continuously using volatile memory accesses, waiting for a change of state before advancing. An SM can never put a block to sleep (in the GPU programming model, a block must execute until completion and cannot be swapped out); at least one thread will always request cycles on the SM\@. And the SM cannot temporarily reallocate resources such as registers and shared memory; the block must fully finish execution first. This is not a sleep in the CPU sense of the word (meaning a thread consumes no processor-specific resources), but it is the least performance-impacting method of waiting available on the GPU\@.

  \item[Fetch-and-Add Mutex] A common instruction on many processors is the ``fetch-and-add''/``atomic increment'' instruction. This instruction can be used to write an efficient mutex with minimal atomics. Essentially the mutex has two variables: a \code{ticket} and a \code{turn}. In \code{lock()}, a thread uses fetch-and-add to atomically increment \code{ticket}, then waits until \code{turn} matches the returned value from the fetch-and-add. In \code{unlock()}, a thread simply uses fetch-and-add to increment \code{turn}.

  \item[Backoff] When a spinning thread does meaningless work to temporarily relieve contention of a resource (e.g.\ the memory bus).

  \item[Centralized Algorithm] When all participating threads/blocks use a single resource, such as the same word in memory, to complete the majority of an operation.

  \item[Decentralized Algorithm] When each participating thread/block uses its own unique resource, such as a distinct word in memory, to complete the majority of an operation.

\end{description}

There are many design decisions we must make for each primitive. Should we use a spin or sleep strategy? Should we use a backoff algorithm? Should we use a centralized algorithm, or a distributed/decentralized algorithm? The lowest common denominator of GPU synchronization primitive is the spin lock, which uses many atomic operations; we believe we can achieve better performance by designing primitives that limit the number of atomic accesses, because atomics are always slower than volatile accesses on both classes of GPU\@.

Backoff will probably not help Tesla algorithms, simply because DRAM operations are so slow, the backoff would have to be very large to compensate, and still might trigger contention due to the number of concurrent accesses. On Fermi, backoff could prove quite useful. The execution time of an atomic is very quick, thus allowing for much smaller windows in the backoff algorithm.

It is hard to use an efficient distributed algorithm for semaphores and mutexes on the GPU, simply because the GPU does not handle linked structures very well (Section~\ref{sec:background}). Thus, we only consider centralized algorithms for both of these primitives. Barriers are an exception though, as every block must participate in every barrier. This allows the use of a faster, decentralized algorithm.

\paragraph*{\textbf{Barrier}}

The barrier is the one synchronization primitive that already has a high-performance implementation on the GPU\@. Xiao and Feng's barrier~\cite{Xiao:2010:IBG} (the ``XF'' barrier) is ideal for the GPU\@. It does not use atomics, has minimal contention in writes (memory lines are bigger than four words), very little contention in reads, and even uses coalescing for both reading and writing. It uses a decentralized, sleeping approach. On top of that, its memory footprint is very small (it uses at most eight times the number of SMs in memory words). The algorithm dictates that each block, upon arriving at the barrier, sets a flag in a specific location---typically in an array of size equal to the number of blocks. A single block (the master block, arbitrarily defined as block~$(0, 0, 0)$) then has all of its threads check the array (each thread checks one or more unique positions) and progress to an intra-block barrier once all other blocks have entered the barrier. Once the threads of the master block pass the intra-block barrier, they change the flags (again, each setting unique positions) to tell the waiting blocks that they may now progress beyond the barrier. Xiao and Feng showed that other barrier methods, specifically a two-stage atomic counter and a hierarchical tree-based barrier, both have worse performance than the XF barrier. However, their results were on Tesla, and we wanted to test our machine abstraction to see how the atomic barrier would perform and scale on Fermi.

%'

This flavor of global barrier requires violating CUDA best practices by scheduling only enough blocks to fully saturate a GPU\@. However, many applications already use this scheduling technique (``persistent threads'') and benefit from it. The barrier is the only primitive with this drawback.

\paragraph*{\textbf{Mutex}}

We use a spin-lock mutex as a baseline for comparison with other mutex implementations. Our spin-lock implementation uses the lock in Algorithm~\ref{listing:spinlock}. Assuming we start with an atomic variable $M$ whose initial value is zero, \code{lock()} simply calls \code{atomicExch($M$, 1)}, which exchanges $M$ with 1 and returns the old value of $M$. If the value returned is 0, then the block now owns the lock, otherwise \code{lock()} loops and tries again. The method for \code{unlock()} is a simple assignment of 0 to $M$. Spin locks have two design flaws though: high contention and heavy use of atomics (every block spins on the same variable).

To mitigate the atomic contention of spin locks, which is bad on Fermi and very bad on Tesla, we added a backoff algorithm. Backoff helps ease contention by executing small sleeps between reads in the same way that a traffic light that gates a freeway entrance helps overall freeway throughput. Due to the uncertainty involved in scheduling, we see some level of atomic contention but still gain sizable speedup and improve performance at scale. On Tesla, backoff is not quite as beneficial as on Fermi because atomic operands come from DRAM and require significant time to fetch. The sleep time necessary to space out reads could very likely negate any benefit backoff might have on Tesla.

We use a small GPU sleep to achieve backoff. Upon entering \code{lock()}\footnote{Backoff in \code{unlock()} is unnecessary, because the block already owns the lock when it enters \code{unlock()}.}, each block maintains an iteration counter $I$, starting with an initial value of $I_\text{min}$ and a maximum value of $I_\text{max}$ (both are configurable in our library at compile time). After each unsuccessful lock attempt, the block sleeps $I$ time units (where $I$ is the time it takes to perform one noncontentious volatile read), then increment $I$. If $I$ is greater than $I_\text{max}$, the block resets $I$ to $I_\text{min}$. Pseudocode for the spin-lock mutex (both with and without backoff) is in the appendix as Algorithm~\ref{listing:spinmutex}.

Backoff reduces/eliminates contention, but it does not meet one of our most important design goals: limiting the number of atomic accesses per operation. One way to limit the necessary number of atomics is to use a method where all atomic operations are done up front, and only volatile memory accesses are required later. The fetch-and-add (FA) mutex algorithm~\cite{Wikipedia:2011:FAA} has the potential to do just that, but requires modifications to make it suitable for the GPU\@. The FA mutex also has a significant advantage over the spin lock on any class of GPUs---it is fair. A spin lock will let in whichever block happens to get lucky. The FA mutex gives access to the critical section to blocks in the order in which they request access.

The standard implementation of FA requires that when a block waits, it ``takes a ticket'' by atomically incrementing a variable, then it ``waits for its turn'' by sleeping on another variable. When a block posts, it simply increments the ``turn'' variable. This method runs well on the GPU, but we improve performance on the GPU by adding backoff to the polling section of \code{lock()}. And in \code{unlock()}, we do not use any atomics. FA has the most potential for Tesla as it uses a sleeping method instead of spinning. On Fermi, sleeping is not much faster so we will not see the same level of gains. Pseudocode for the FA mutex is in the appendix under Algorithm~\ref{listing:famutex}.

As a design alternative, we also explored a ring-buffer based sleeping mutex. As a block arrived, it would place itself at the end of the ring buffer and constantly check to see if the item at the front was itself. To unlock the mutex, a block would simply increment the head pointer. On the GPU, the algorithm was inferior in all aspects (more memory consumption, more reads per attempt, same amount of contention) to the FA mutex. We can easily explain this behavior with our GPU abstraction. The ring-buffer mutex has roughly the same amount of atomic contention as FA, but twice as many reads. And on Fermi, the head-pointer read will be serialized if at any time, one block posts while many blocks are waiting.

On the GPU, it is not necessary, or even possible, to implement the typical Linux-style mutex~\cite{Russell:2002:FFF}. The mutex consists of an aggressive spin lock followed by a blocking lock, and the GPU does not allow a thread to block. To achieve similar (but not identical) behavior, we could use an aggressive spin lock that eventually reverts to using backoff.

\paragraph*{\textbf{Semaphore}}

We again used a spin-lock as our baseline for semaphores. We modify the spin-lock algorithm slightly to compensate for the lack of generic atomic transactions on the GPU---specifically, we need a ``perform OP if greater than zero'', but the closest operation on NVIDIA GPUs is ``swap if equal to''. For a semaphore with an initial count of $X$, we initialize an atomic variable $S$ to $X + 1$. In \code{wait()}, a block will loop and call \code{atomicExch($S$, 0)}. If the value $V$ returned is zero, then another block has the lock and this block simply loops. If $V$ is one, which is $(X + 1) - X$, it means that the semaphore is at capacity and we simply set $S$ back to one. If $V$ is greater than one, the block has control of the lock and the semaphore is not at capacity. The block will then set $S$ to $V - 1$ and return from wait. In \code{post()}, the algorithm is similar. The block will keep trying to acquire the lock by calling \code{atomicExch($S$, 0)}. If the returned value $V$ is not zero, then the block has the lock and sets $S$ to $V + 1$, then exits. This implementation has more reads and writes than a spin-lock mutex, simply because of the extra checks involved in a semaphore, and because the atomic operations on the GPU are more restrictive than those offered on a modern CPU\@. The spin-lock semaphore has three drawbacks: even in \code{post()}, a block may have to spin; both \code{wait()} and \code{post()} have heavy atomic contention; and multiple accesses are required to both lock and unlock the semaphore.

To mitigate the contention inherent in the spin-lock semaphore, we explored a backoff implementation similar to our mutex backoff. We did not introduce backoff into \code{post()} because it should proceed quickly and aggressively, as usually many more blocks are waiting than posting. Since our spin-lock semaphore uses a locking mechanism, the backoff in \code{wait()} eases contention, both in \code{wait()} and \code{post()}. Pseudocode for the spin-lock semaphore (both with and without backoff) is in the appendix under Algorithm~\ref{listing:spinsem}.

Just as with mutexes, neither the spin-lock nor spin-lock-with-backoff addresses our biggest goal, to bound the number of atomic accesses necessary for an operation. We return to the FA implementation of a semaphore and adapt the algorithm to the GPU\@. Given a semaphore $S$ with initial value $V$, we initialize an atomic variable $C$ (count) to zero. When a block calls \code{wait()}, it atomically increments $C$ and retrieves the old value of $C$. If the old value of $C$ is less than $V$, then the block proceeds into the critical section. Otherwise, it uses the FA model and takes a ticket. When waiting for the ticket to be called, the block will check if the current turn is greater than or equal to its ticket number. Once this condition is met, the block will proceed into the critical section. When a block calls \code{post()}, it first decrements $C$. If the previous value is greater than $V$, then the block will increment the turn counter to let a new block into the critical section. Otherwise, it will simply exit. This implementation guarantees that only one or two accesses per block happen in \code{wait()} and \code{post()}, something that a spin-lock semaphore can never guarantee. Pseudocode for the sleeping semaphore is in the appendix under Algorithm~\ref{listing:sleepsem}.

By incrementing the turn counter only when a block is waiting, we ensure that we do not violate the semaphore count. And by using an FA-style approach, we assure fairness in that blocks will be allowed to proceed in the order in which they arrive.

The modified FA implementation we wrote has the main advantage of FA in that the algorithm performs all of the necessary atomic accesses as early as possible and it executes only a finite number of atomics. It also has an advantage over FA in that if the semaphore is under capacity, it requires only one single memory access (an atomic increment), which means it does not require any spinning or sleeping.

\paragraph*{\textbf{Summary}}

We developed several techniques to achieve high synchronization primitive  performance on GPUs when compared to baseline spin-lock primitives:

\begin{itemize}

\item \textbf{Backoff} eases atomic pressure and allows the atomic units to flush their queues on Fermi, thus turning a spinning algorithm into a sleeping algorithm.

\item By \textbf{replacing as many atomic accesses as possible with non-atomic accesses}, we make each primitive much more fast and efficient.

\item \textbf{Bounding the number of atomics} ensures that, given that an atomic unit has enough time to flush its queue, we will achieve the lowest wait time possible.

\item We \textbf{ensure fairness}, something a spin lock does not do, by employing algorithms that guarantee threads gain access in the order in which they arrive.

\end{itemize}

\paragraph*{\textbf{API}}

We firmly believe that it is important to present a unified API across all GPUs, one where users have access to all implementations, and the default is the most high-performance implementation for their platform. We provide such a library (we believe it is the first ever for the GPU), and list the functionality of each primitive in Table~\ref{table:primlisting}. This library is available as open-source for download at \textit{Google-Code Link Will Be Given in Published Paper}.

\begin{table}
  \noindent
  \begin{center}
    \begin{small}
      \begin{tabular}[th]{lp{2in}}
        \toprule
        Primitive : Function    & Description \\
        \midrule
        Barrier   : Barrier()   & Forces all blocks to wait within this function until every block has entered this function. \vspace{5pt} \\
        Mutex     : Lock()      & Attempts to acquire a mutually-exclusive lock. If the mutex is already locked, a calling block will wait until it acquires the \vspace{5pt} lock. \\
        Mutex     : Unlock()    & Releases the mutex. \vspace{5pt} \\
        Semaphore : Wait()      & Attempts to acquire a slot within the semaphore. If the semaphore is at capacity, the calling block will wait until a spot opens and it can acquire a spot. \vspace{5pt} \\
        Semaphore : Post()      & Releases one slot within the semaphore. \vspace{5pt} \\
        \bottomrule
      \end{tabular}
    \end{small}
    \caption{API Listing: Functions provided for each primitive, and its effect. Note that we exclude non-blocking options since certain implementations (e.g.\ FA mutex) do not support such behavior.\label{table:primlisting}}
   \end{center}
\end{table}

%% file: results.tex
For all of our tests, we used synchronization-primitive operations per second as the primary figure of merit. For barriers, all blocks must complete the barrier for a single operation. For mutexes and semaphores, only one block must complete a lock/wait per operation. Each test is comprised of a single kernel that uses one unrolled loop to execute 1000 operations. For mutexes, this means a lock immediately followed by an unlock. For semaphores, this means a wait followed immediately by a post. For barriers, it simply means executing a global barrier.

\paragraph*{\textbf{Testing Methodology}}

For both Tesla and Fermi, we used machines with a quad-core AMD Opteron 2216 with 8 GB of RAM running 64-bit Rocks Linux and the 2.6.32 kernel. Our Tesla card was a GTX295 and our Fermi card was a GTX580. The GTX295 has 30 SMs, which can each handle a maximum of 8 blocks simultaneously, so we run tests from 1 to 240 blocks. The GTX580 has 16 SMs, which can each handle a maximum of 8 blocks simultaneously, so we run tests from 1 to 128 blocks. On each GPU, we use 128 threads per block. For each test we perform, each block performs 1000 instances of the operation. We chose 1000 operations per block as at that number, the trends were smooth, and as we increased the number of operations from there, the trends did not change.

It is not important to compare the difference between performance on Tesla with 240 blocks and Fermi with 128 blocks. We expect different performance trends on the GPUs due to their memory system. The important things to compare are performance of each primitive implementation on the same GPU, and how the implementation scales from one block to the maximum number of blocks on each GPU\@. The most interesting comparison across GPUs is simply that a particular primitive implementation may be the best on one GPU but not on another.

It is important to restate this here: for mutexes and semaphores, even though we use 128 threads per block, only one thread in each block accesses the primitive. In the event of multiple resources accessed per block, it is entirely possible for more than one thread per block to try to lock different resources. However, if more than one thread in a block wants access to the same resource, it is much more efficient to use a block-level scheme (e.g.\ a reduction) and then have one thread access the primitive.

\paragraph*{\textbf{Barrier}}

\begin{figure*}
  \begin{tabular}{cc}
    \includegraphics[width=3.25in]{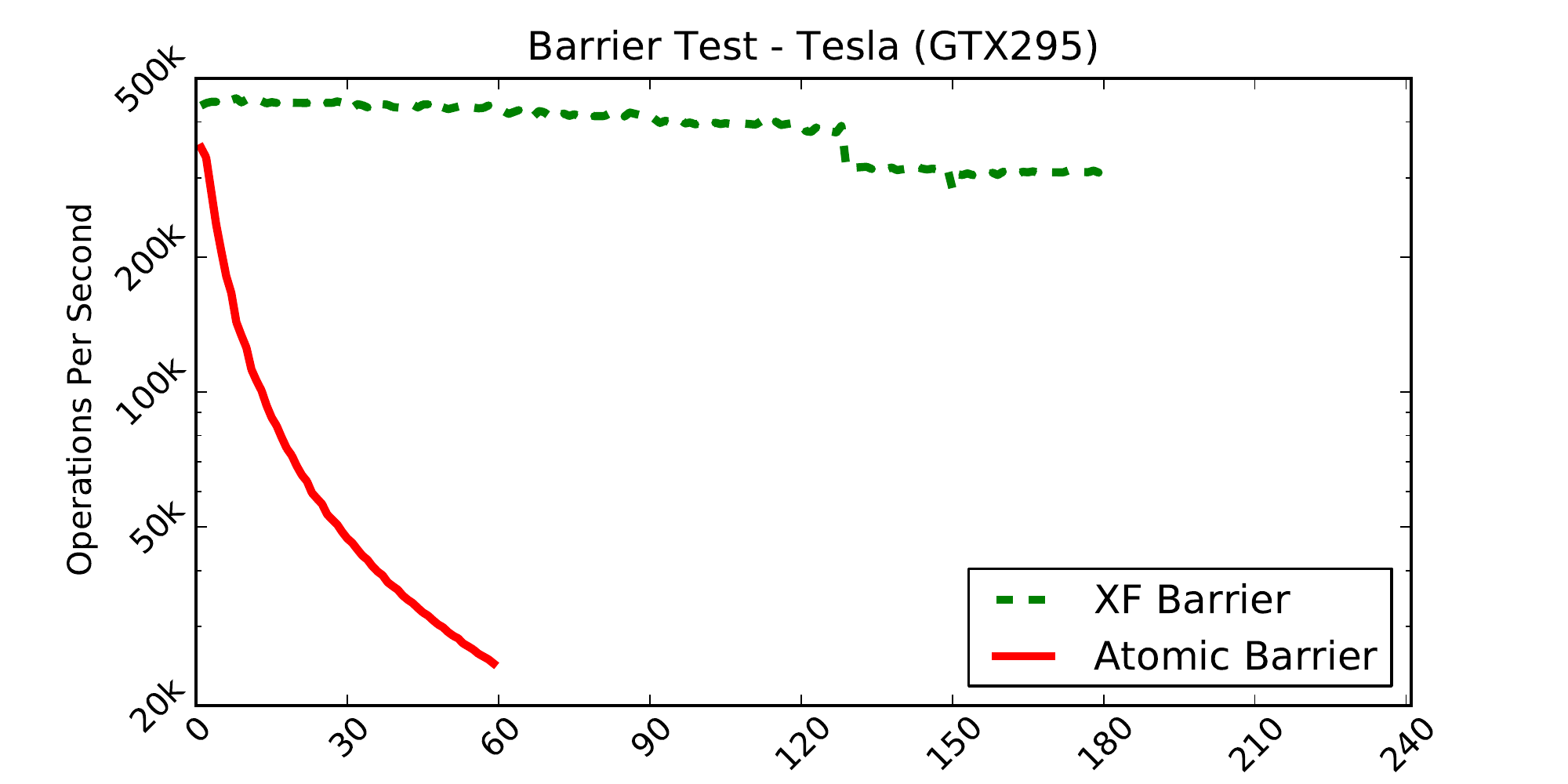} &
    \includegraphics[width=3.25in]{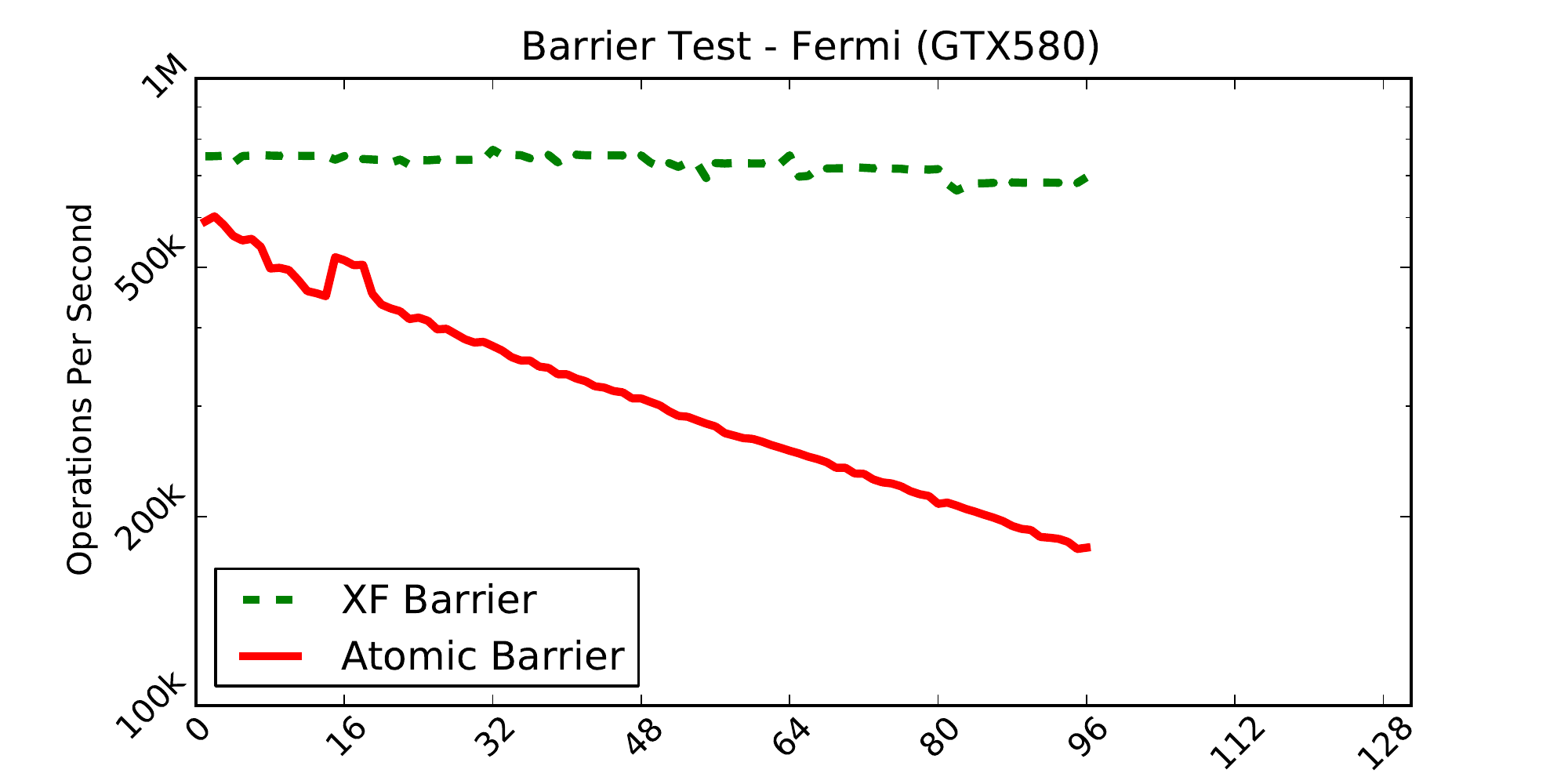}\\
    \scriptsize{\textsf{\# Blocks}} & \scriptsize{\textsf{\# Blocks}}
  \end{tabular}
\caption{Barrier Results: The x-axis is the number of blocks used by the kernel. The y-axis is on a log scale. Higher y-values are better. The XF barrier is always faster than the two-stage atomic barrier. On Tesla, the atomic barrier starts to drastically under perform after 2 blocks per SM, hence why we cut off any performance measurements at that point. The XF barrier requires more registers, allowing us to only time with 6 blocks per SM instead of 8. The dip in performance in the XF barrier on Tesla is because we use 128 threads per block, thus when the number of blocks goes above the number of threads, each thread must do extra work. The sudden performance jump of the atomic barrier on Fermi is due to a peculiarity of the atomic unit, and how it handles releasing and then reacquiring a line (it must perform thrice as many tag lookups in this case), as opposed to the behavior when just holding a line. \label{fig:barrier_results}}
\end{figure*}

On Tesla, the performance of the atomic barrier degrades rapidly as we add blocks, thus we do not show trends beyond 60 blocks. The XF barrier requires many registers to complete, so we could only run up to 6 blocks per SM\footnote{This is an important point, as using the XF barrier will limit resource usage in the rest of the kernel.}. On Fermi, the atomic barrier performs much better and is more scalable than on Tesla. We present the full results for both Tesla and Fermi in Figure~\ref{fig:barrier_results}. On Tesla, though both barriers start out at approximately the same rate, the atomic barrier rapidly declines and falls below 5\% of the XF barrier at sixty blocks. On Fermi, the atomic barrier starts at about 75\% of the XF barrier, but drops in performance all the way down to roughly 30\% at full scale.

\paragraph*{\textbf{Mutex}}

\begin{figure*}
  \begin{tabular}{cc}
    \includegraphics[width=3.25in]{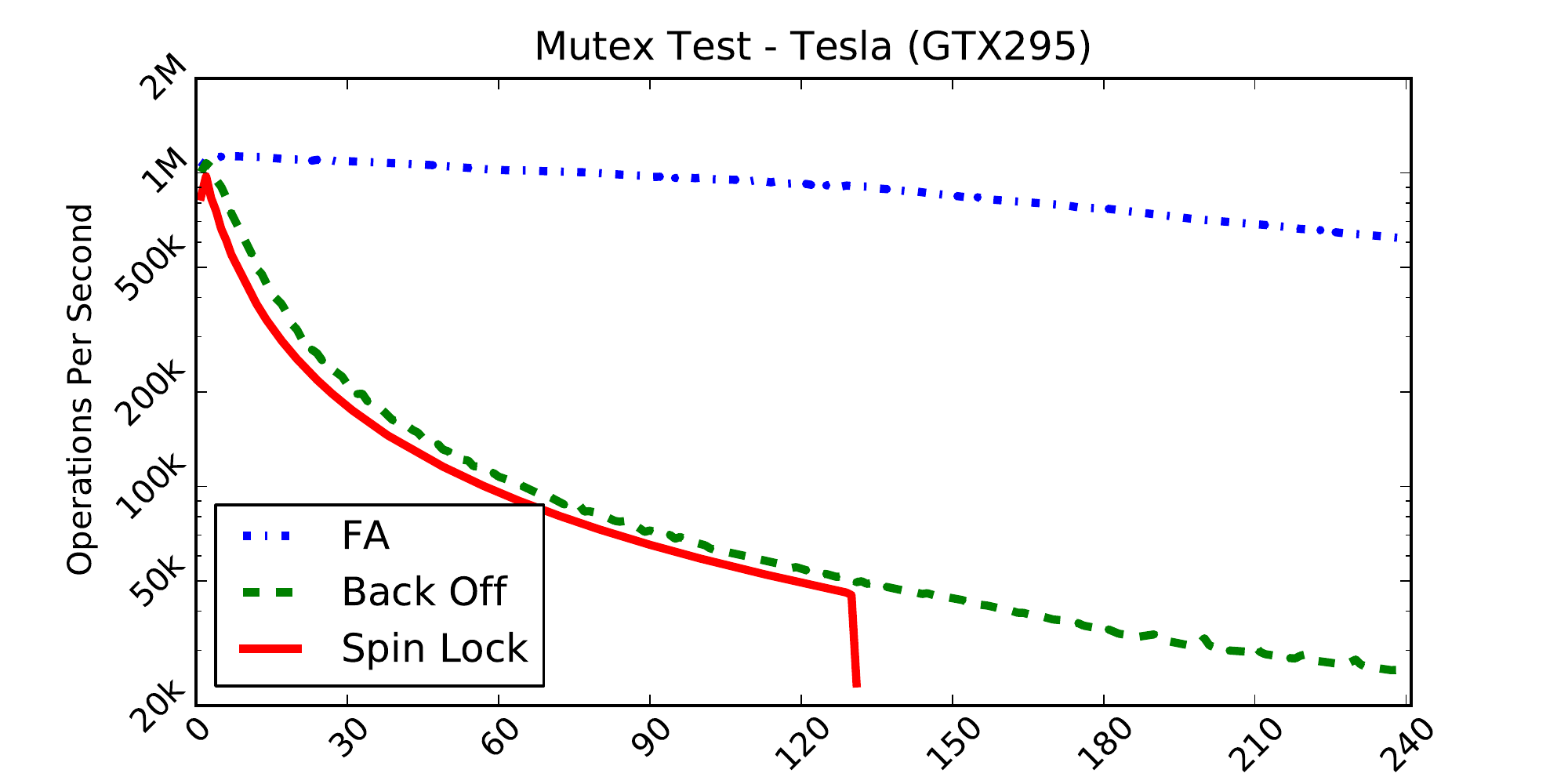} &
    \includegraphics[width=3.25in]{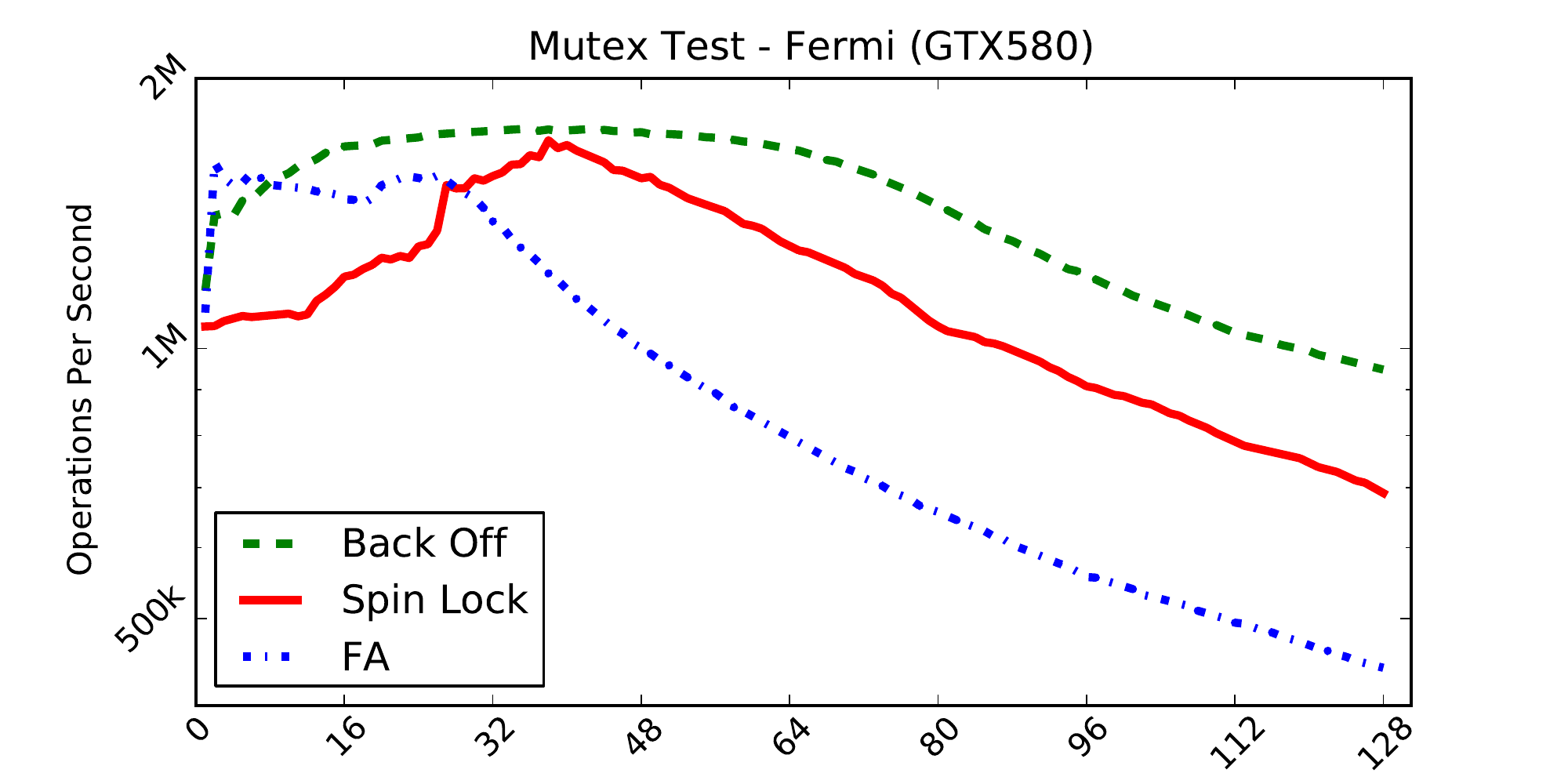}\\
    \scriptsize{\textsf{\# Blocks}} & \scriptsize{\textsf{\# Blocks}}
  \end{tabular}
\caption{Mutex Results: The x-axis is the number of blocks used by the kernel. The y-axis is on a log scale. Higher y-values are better. The spin lock on Tesla suffers from unpredictable performance (the performance is both unpredictable and poor) after approximately 130 blocks. On Fermi, the spin lock performance is stable and fast enough to keep improving the operations per second up to approximately 40 blocks, at which point the atomic unit becomes the bottleneck and the speed at which it can process slows. On Tesla, backoff helps the spin lock in terms of absolute performance, but does not improve scalability. On Fermi, backoff actually does both; it helps give better and more predictable performance, and it gives better scalability. On Fermi, FA gives poor performance compared to the spin lock because of the number of extra accesses required and the fact that the atomic unit holds a memory line until its queue is flushed, which serializes atomic accesses. On Tesla however, FA outperforms the spin lock because the speed of contentious atomics is so much slower than that of contentious volatile accesses. \label{fig:mutex_results}}
\end{figure*}

The spin-lock mutex performance degrades rapidly on Tesla once we pass a certain threshold (times are unpredictable and poor at/after approximately 130 blocks), thus we do not present timings past that point. The three other mutex implementations scale well enough that we can test up to 240 blocks. On Fermi, all mutex implementations perform well enough to collect results with up to 128 blocks simultaneously locking and unlocking. The full results are shown in Figure~\ref{fig:mutex_results}.

Backoff helps only slightly on Tesla, giving a barely noticeable performance boost. It does, however, ensure smooth scaling (even if the scaling is negative). At full scale on Tesla, a spin-lock mutex with backoff runs at less than 5\% the speed of a sleeping mutex. On Fermi, though, backoff helps immensely. Not only does it give a much smoother scaling trend, but at scale it yields an almost 45\% speed gain over the next fastest implementation (the plain spin lock). The sleeping mutex on Fermi is slow, coming in at roughly half the speed of a spin lock with backoff.

\paragraph*{\textbf{Semaphore}}

\begin{figure*}
  \begin{tabular}{cc}
    \includegraphics[width=3.25in]{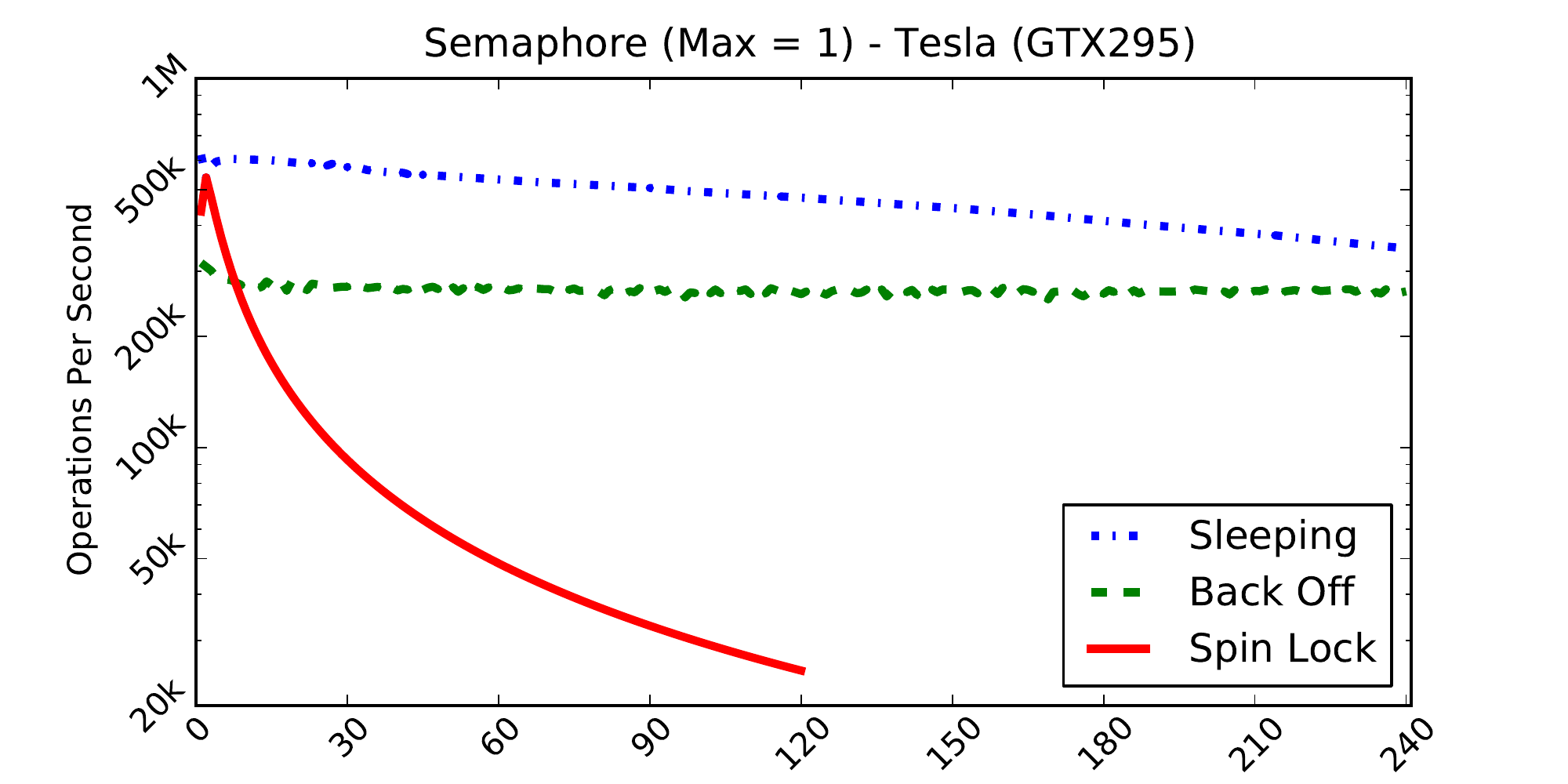} &
    \includegraphics[width=3.25in]{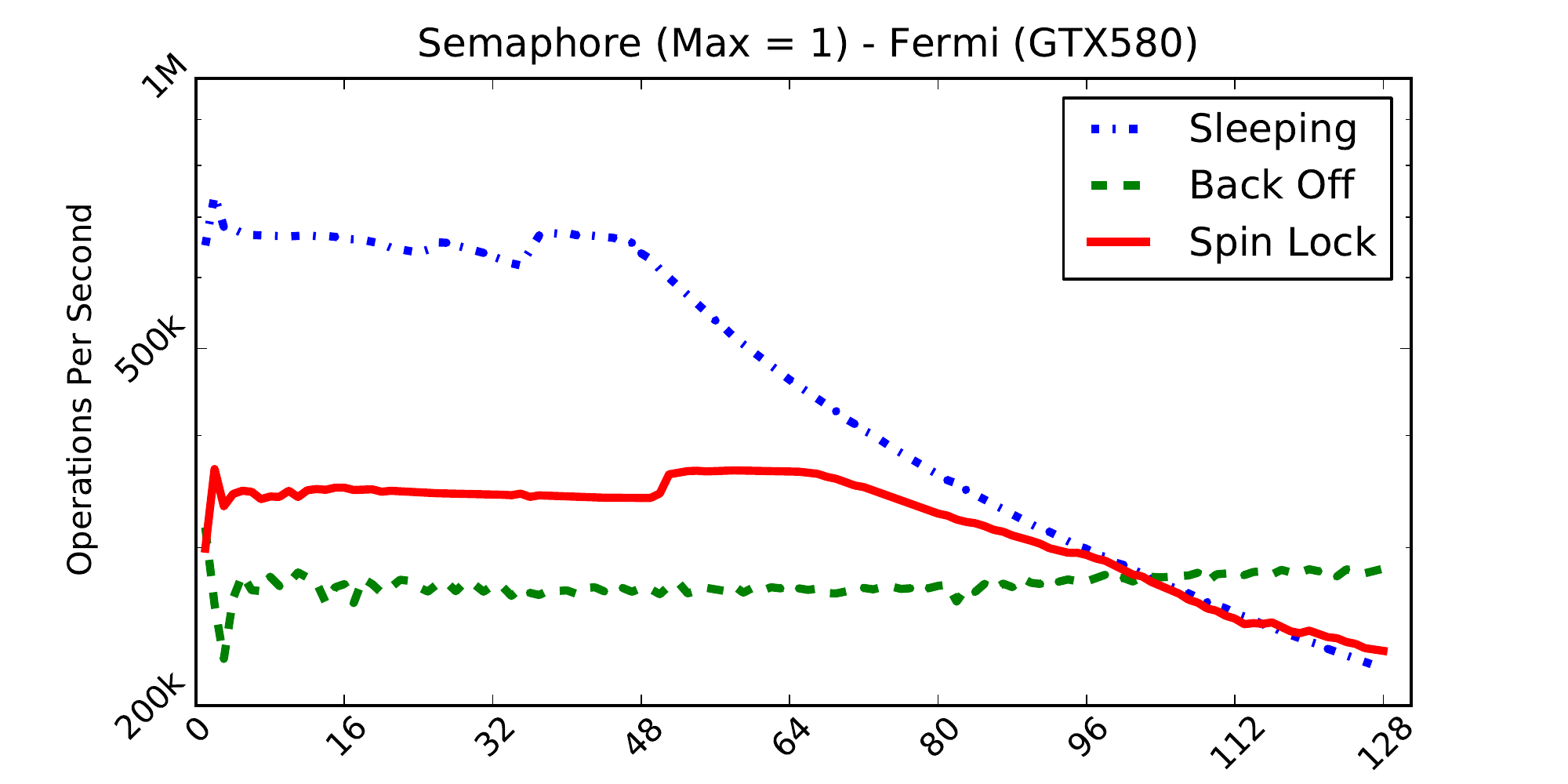} \\
    \scriptsize{\textsf{\# Blocks}} & \scriptsize{\textsf{\# Blocks}} \\

    \includegraphics[width=3.25in]{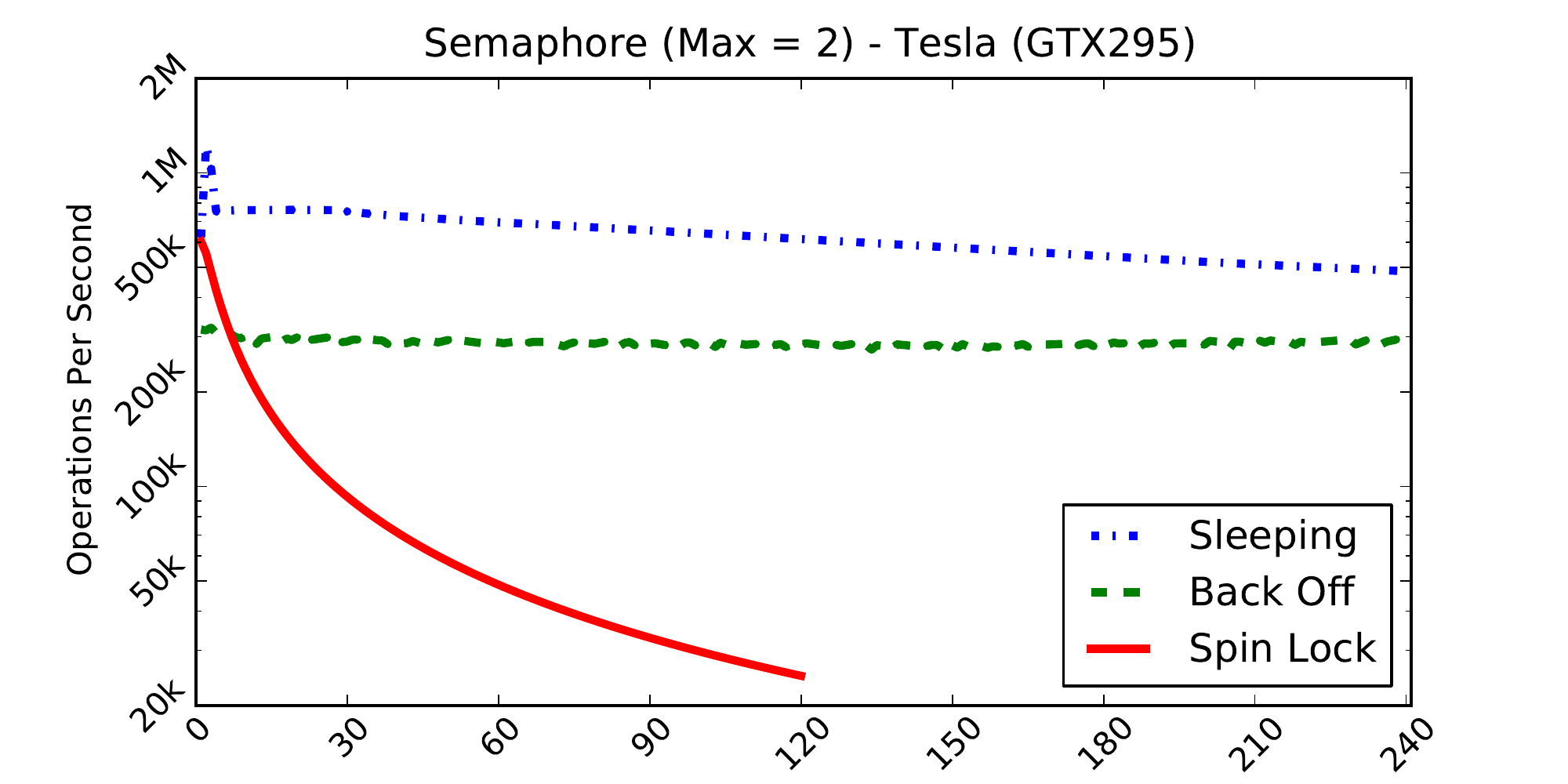} &
    \includegraphics[width=3.25in]{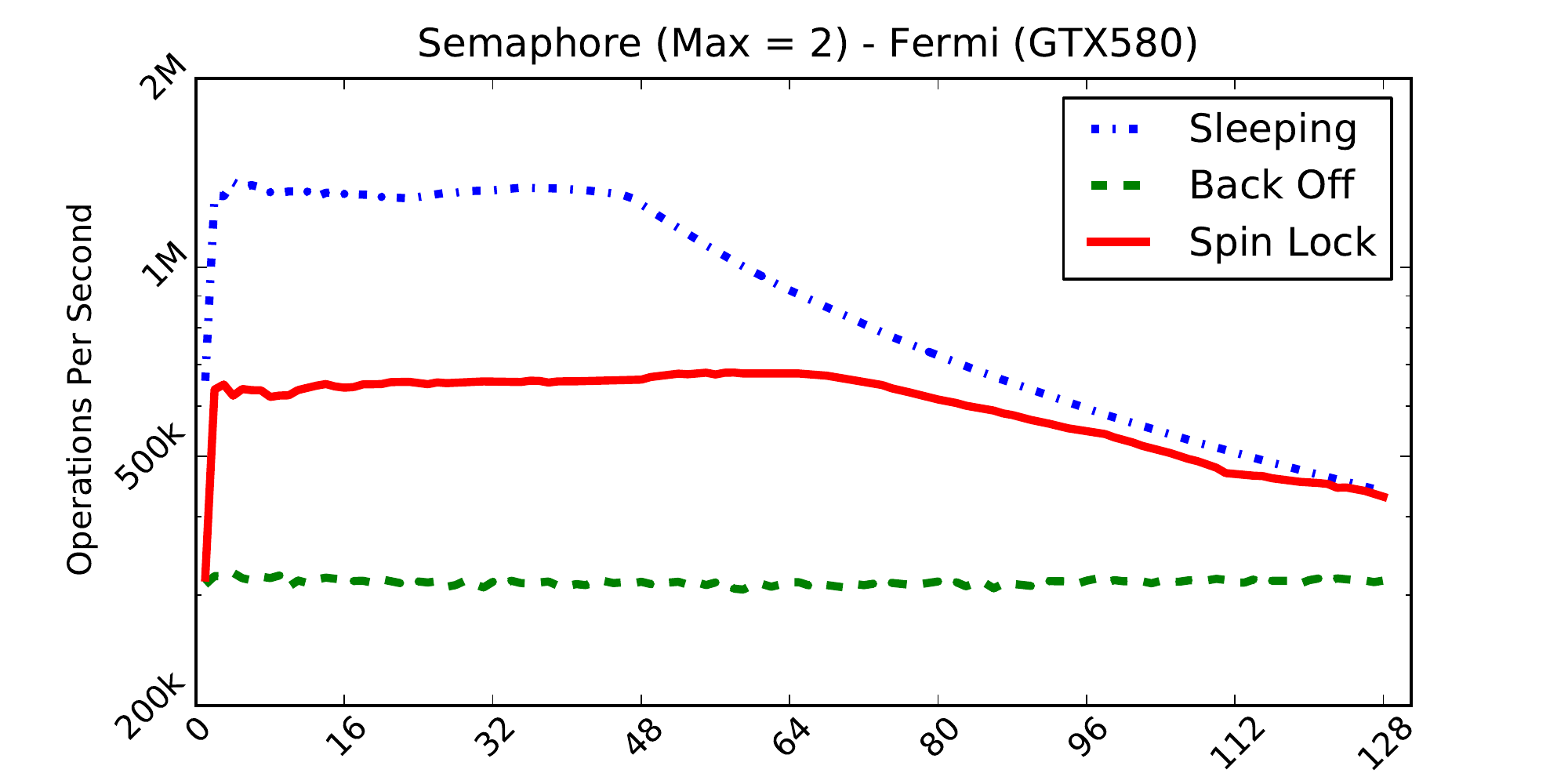} \\
    \scriptsize{\textsf{\# Blocks}} & \scriptsize{\textsf{\# Blocks}} \\

    \includegraphics[width=3.25in]{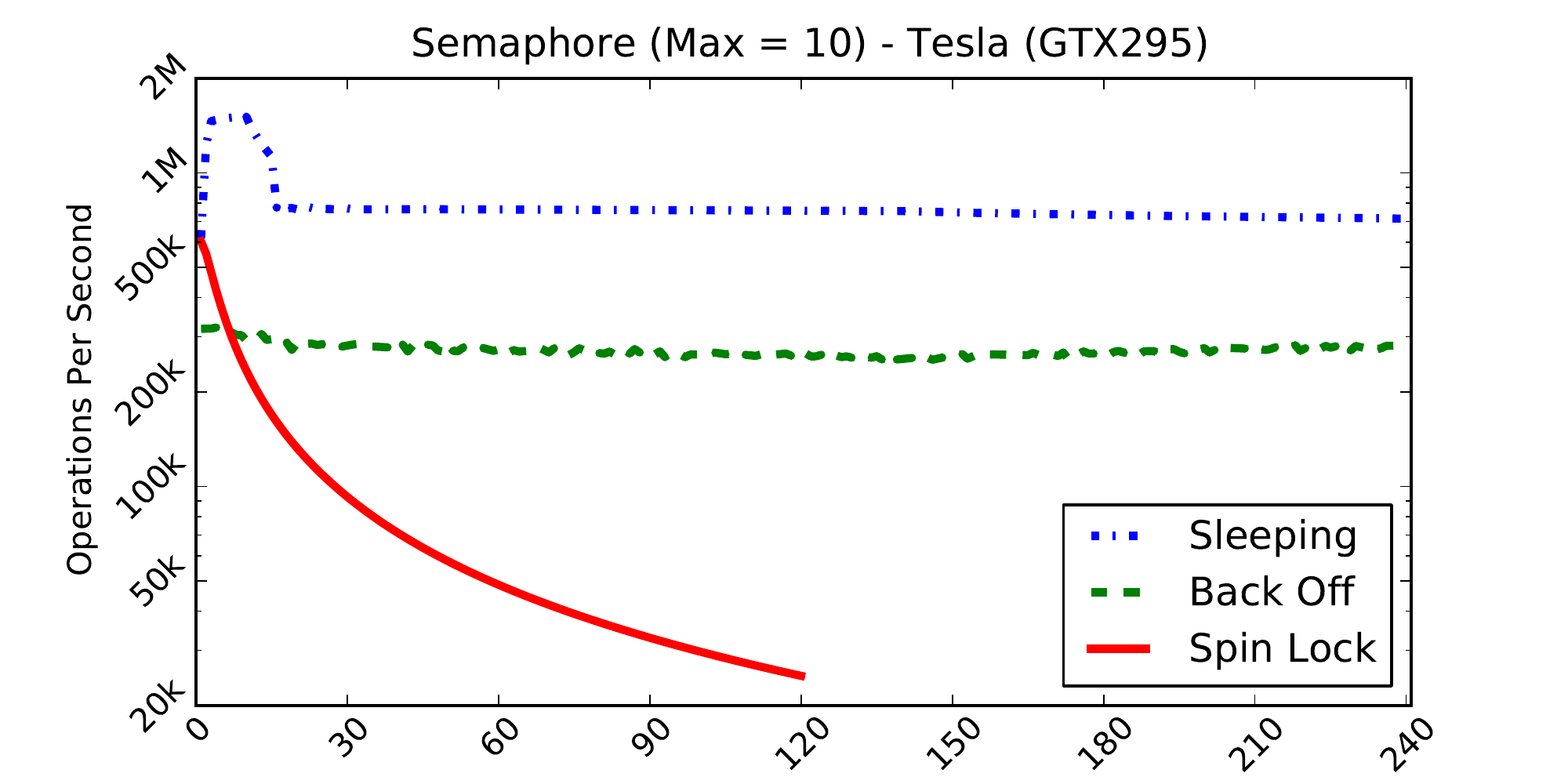} &
    \includegraphics[width=3.25in]{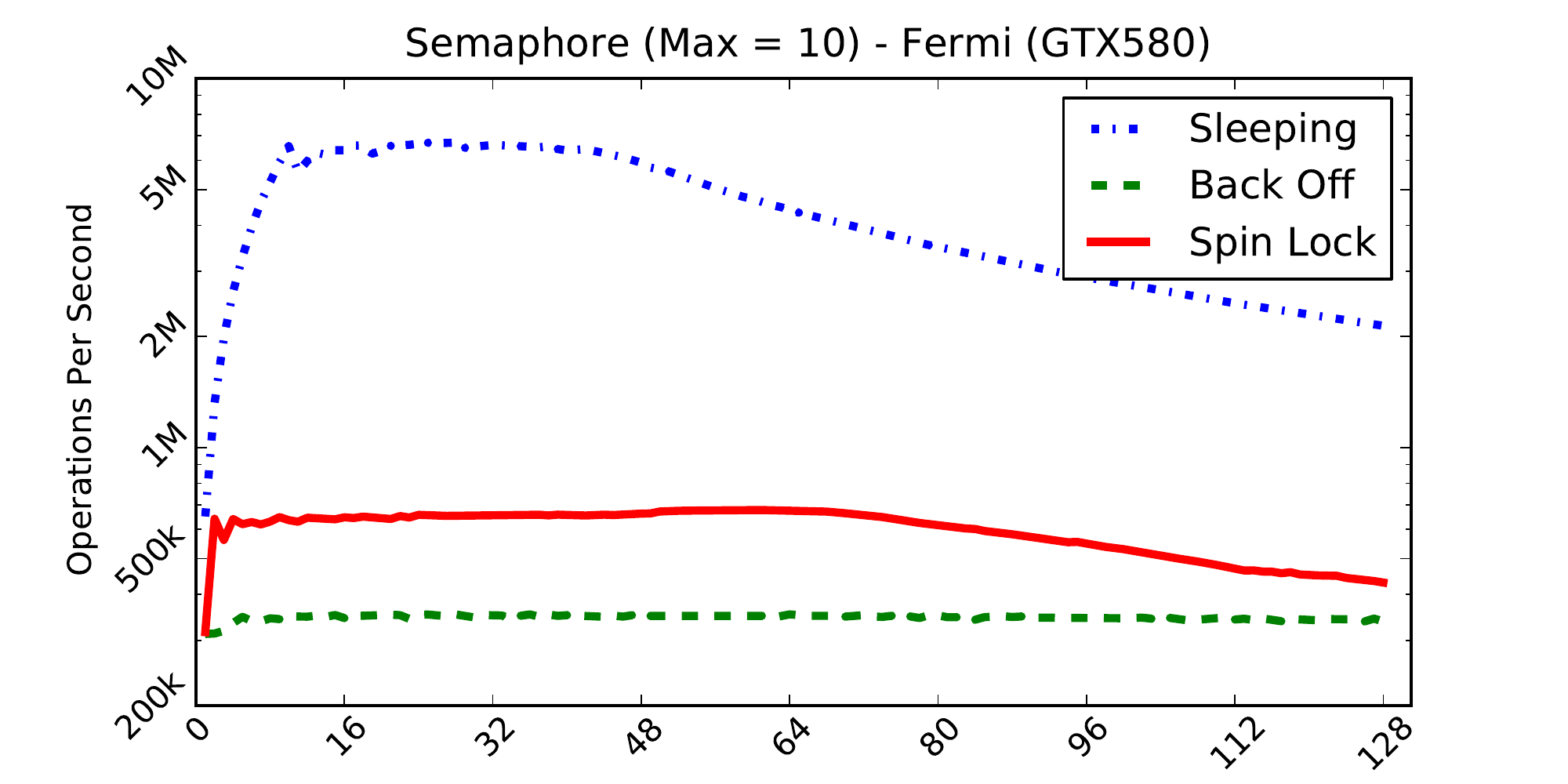} \\
    \scriptsize{\textsf{\# Blocks}} & \scriptsize{\textsf{\# Blocks}} \\

    \includegraphics[width=3.25in]{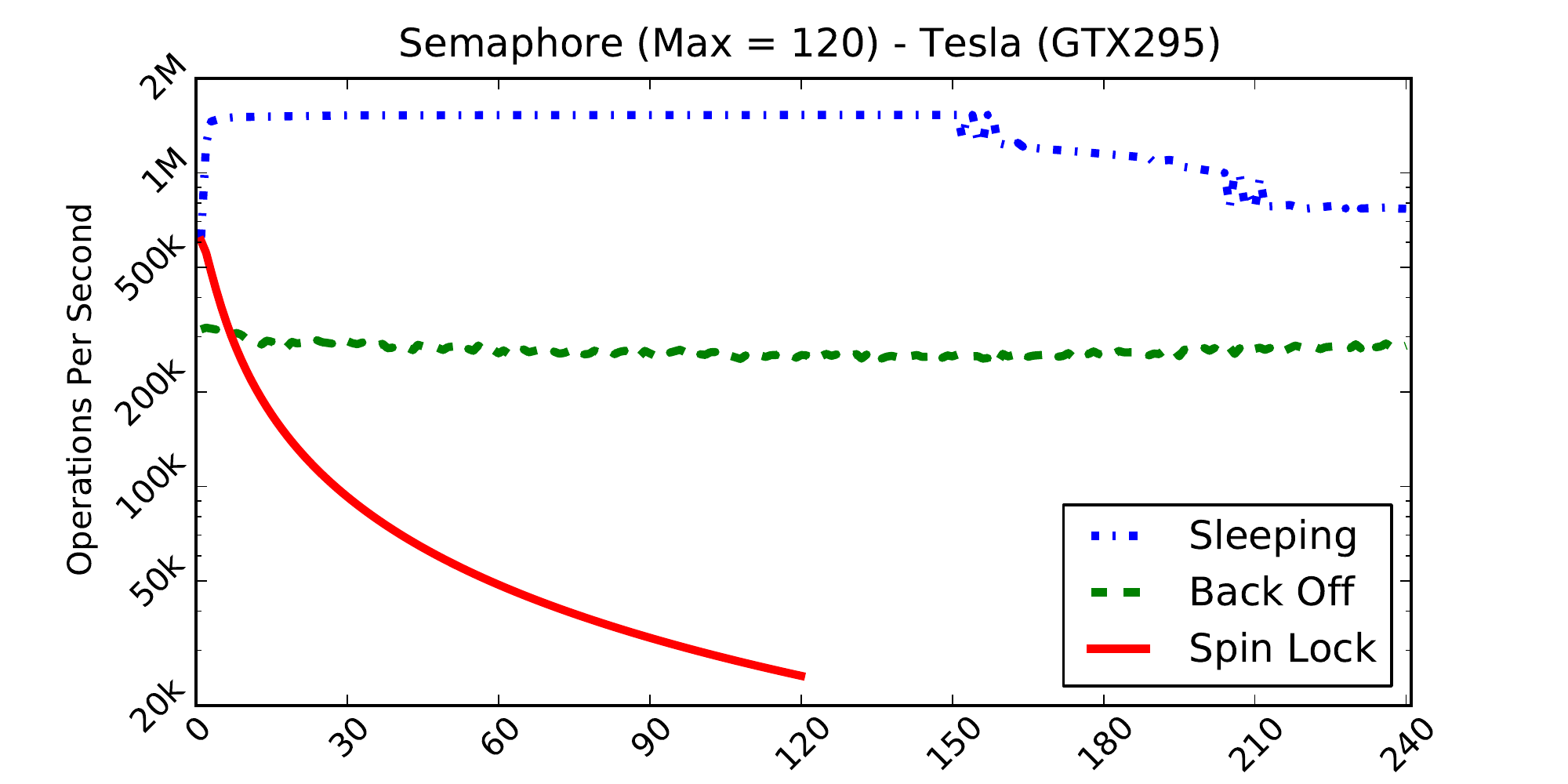} &
    \includegraphics[width=3.25in]{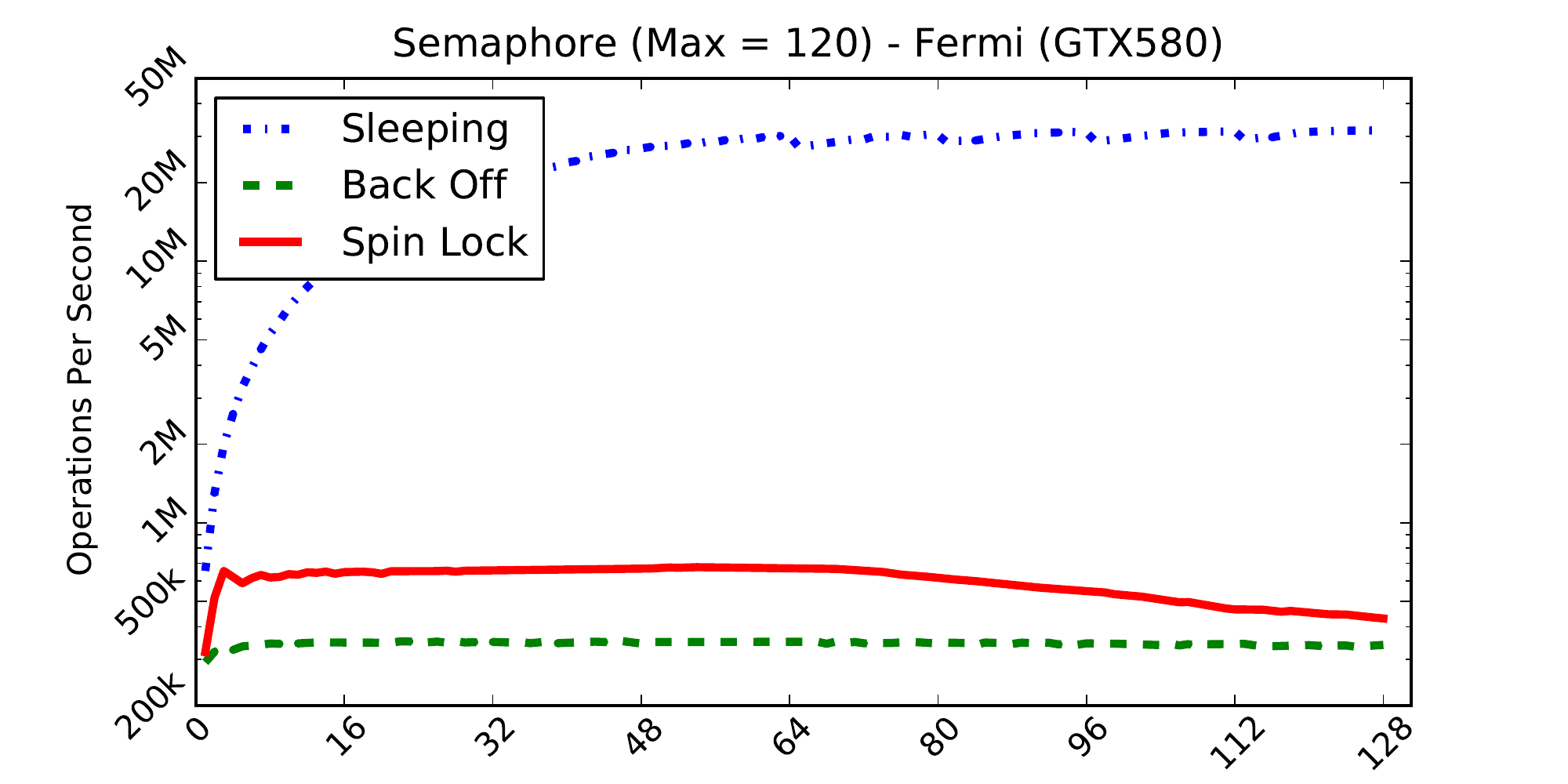} \\
    \scriptsize{\textsf{\# Blocks}} & \scriptsize{\textsf{\# Blocks}} \\
  \end{tabular}
  \caption{Semaphore Results: The x-axis is the number of blocks used by the kernel. The y-axis is on a log scale and the extents differ between graphs. Higher y-values are better. The spin-lock semaphore on Tesla requires such a long time for \code{wait()} that its behavior seems invariant of the initial semaphore value. On Fermi, the behavior improves as the initial value increases, but not by very much. Backoff helps significantly on Tesla as it allows the aggressive \code{post()} operation to complete quickly. On Fermi, backoff does not relieve enough contention to make \code{post()} any faster, and in fact tends to make the algorithm slower than the baseline in most cases. The sleeping algorithm is almost always the fastest implementation, and the one with the best scalability, both on Tesla and on Fermi, with every range of initial values. This is because \code{post()} on a sleeping semaphore does not have any spinning and requires at most two atomic operations, and \code{wait()} on a sleeping semaphore not yet at capacity simply requires a single atomic and no spinning. The one case where a sleeping semaphore is slower than a spinning semaphore is when it is used as a mutex on Fermi. This is most likely due to the high contention of several atomic variables, compared to the high contention of the single variable used by the spin lock. \label{fig:semaphore_results}}
\end{figure*}

We tested the semaphore not only with all possible numbers of blocks, but also with a varying initial value (the maximum number of concurrent blocks the semaphore allows in the critical section). From our test group, we chose four specific values that exemplified the trends: 1, 2, 10, and 120. The scalability of the spin-lock semaphore on Tesla is poor after 120 blocks (4 blocks per SM), thus we do not show results for the spin-lock semaphore beyond that. The results for all eight combinations of GPUs and initial values are shown in Figure~\ref{fig:semaphore_results}. On both Tesla and Fermi, the sleeping semaphore is generally the fastest (though with an initial value of 1, on Fermi, the spin-lock semaphore with backoff does overtake it). The sleeping semaphore, most likely due to its fast access when under capacity, scales very well, even with a low initial value. For example, on Fermi, at full scale, the sleeping semaphore with an initial value of 2, 10, and 120 is, respectively, the same speed, $6\times$ faster, and $60\times$ faster than a spin-lock semaphore. On Tesla, we can only compare against the spin lock with backoff. At full scale, with initial values of 1, 2, 10, and 120, the sleeping semaphore achieves performance gains of $1.3\times$, $1.7\times$, $2.5\times$, and $2.7\times$ respectively over the spin-lock semaphore.

While it would be desirable to make many broad conclusions about the GPU in general, each GPU has different characteristics and we cannot say that there is a ``one size fits all'' strategy for every GPU and primitive. Even on the same GPU, with one primitive spinning works well, whereas with another primitive, sleeping works better. Table~\ref{table:bestforgpu} summarizes the best primitive implementation for each GPU\@.

\begin{table*}
  \noindent
  \begin{center}
    \begin{small}
      \begin{tabular}[th]{lllll}
        \toprule
                                                      & Tesla     & Fermi \\
        \midrule
        Best Barrier at scale                         & XF        & XF \\
        Best Mutex at scale                           & FA        & Spin with Backoff \\
        Best Semaphore at scale (low initial value)   & Sleeping  & Spin with Backoff \\
        Best Semaphore at scale (high initial value)  & Sleeping  & Sleeping  \\
        \bottomrule
      \end{tabular}
    \end{small}
  \caption{The best implementation for each primitive, per GPU\@. The XF Barrier is the fastest on both GPUs. The FA mutex is always the best choice on Tesla, while on Fermi it is good for a small number of blocks but the spin lock with backoff quickly overtakes it. The sleeping semaphore is by far the best choice on Tesla in all cases, and almost all cases on Fermi. The one exception is with an initial value of one at high scale. \label{table:bestforgpu}}
   \end{center}
\end{table*}

For example, on Fermi, a spinning semaphore is noticeably slower than a sleeping semaphore, but a spinning mutex (even without any backoff) is noticeably faster than a sleeping mutex, and our machine abstraction helps explain this. With a spinning semaphore, a block must perform multiple atomic operations to wait, and also must spin to post. With a sleeping semaphore, a successful wait may require only a single atomic operation (and at most two atomic operation with a volatile spin), and a post never requires more than two atomic operations and never requires any sort of waiting (whether it be spinning or sleeping). The opposite is true for mutexes; a spinning mutex will outperform a sleeping mutex due to the eccentricities of the atomic pipeline on Fermi and how an atomic unit will hold a memory line hostage under contention, thus making the most important aspect of the mutex implementation the total number of operations (both atomic and volatile) necessary to accomplish a task.

On Tesla, sleeping is always the best performing option. Again, our machine abstraction for Tesla explains this well. The cost of contentious atomics is so high on Tesla, and the atomic unit does not have the same eccentricities as those on Fermi, that performing all necessary atomics up front and then sleeping via VL polling will always outperform spinning via atomics.

On both Tesla and Fermi, backoff with the proper arguments (as outlined in Section~\ref{sec:design}) causes some implementations to perform faster. On Tesla, the speedup ranges from subtle with mutexes (5--10\% speedup) to drastic with semaphores (total change of behavior, much more scalable, more than 10x speedup at scale). On Fermi, the speedup for a spin-lock mutex with backoff is about 60\% at scale. The behavior is also significantly more predictable. Interestingly enough, backoff with semaphores actually causes a drop in performance on Fermi, due to the necessity of multiple atomic operations to wait on a semaphore, combined with the aggressive round-robin scheduling of the memory crossbar on Fermi. Another interesting point about semaphores with backoff is that their performance is largely constant as the number of blocks vary, regardless of the maximum count. One possible explanation of this is that we have an empty critical section and the backoff arguments are such that no thread ever has to wait enter.

With all SMs fully saturated on Fermi, the performance of the sleeping semaphore scales almost linearly with respect to the maximum value. We believe the \code{wait()} is efficient enough with an empty critical section to essentially never require a thread to sleep.

Of course, all of these observations are most noticeable under strong contention, which is the scenario in which a high-performance synchronization primitive is most necessary. Without contention, the cost of each primitive may be modest compared to the total amount of work done in the kernel.

%% file: conclusion.tex
The performance of our algorithms compares favorably to the current baseline in GPU programming. Our FA mutex is almost $40\times$ faster than a spin lock on Tesla, while on Fermi adding backoff to the spin lock gives it nearly a 40\% speedup. Our sleeping semaphore on Tesla is more than $3\times$ faster than the spin lock semaphore, and on Fermi the sleeping semaphore is more than $70\times$ faster than the spin lock semaphore. The primitives we created apply to all GPUs that support atomics. The benchmarks and tests allow us to further hypothesize on what sort of performance we can expect from future GPUs. For example, if atomics someday happen to be as fast as volatile loads and stores, the only important factor would be contention. Thus, a spin lock with backoff might very well be the most high-performance implementation of any primitive. If contention causes worse behavior in a new GPU, backoff might be the most important factor in speeding up a primitive.

We believe it is vital to provide these primitives in a library that is easy for developers to use. We created an API that we feel is well-suited for GPU programmers for many reasons: it is high level, flexible, and easily allows users to change the implementations of their primitives without changing any of their application code. It provides users already experienced with CPU primitives familiar semantics such as a mutex data structure with \code{lock()} and \code{unlock()} functions. The API allows for specificity (e.g.\ a user can specifically create a spin-lock semaphore), but we can also easily provide the most high-performance implementation as a default to users based on their GPU type, allowing them to avoid the lower-level details.

Our tests and benchmarks allow us to predict performance of our implementations on future GPUs. For example, the transition from Tesla to Fermi brought an order-of-magnitude speedup in atomic:volatile speed. The end goal of the hardware vendors is to make atomics as fast as regular loads. This is a hard task for many reasons (extra tag lookups, serialization, line captures, etc.), so while we may not see an order-of-magnitude jump again, we do expect the gap to continue to shrink. As atomics perform increasingly close to volatile accesses, backoff will play an even more important factor. We say this because the contentious:noncontentious ratio on Fermi is around 10:1, which severely impacts the performance of any spinning implementation.

The improvement of atomics actually highlights another area that could prove quite valuable, making contentious accesses faster. An atomic unit on Fermi holds memory lines hostage until it flushes its queue. This also means that it serializes volatile accesses, essentially turning them into atomic accesses. If the atomic unit is able to keep information on whether or not the access was originally atomic, it could potentially flush large pieces of the queue at once. For example, if the next $n$ operations are all loads that were serialized and turned into atomic operations, the atomic unit could handle all $n$ operations at once.

Another potential improvement to the hardware is a global barrier (e.g.\ \code{\_\_syncblocks()}). This is not an inherently difficult operation to implement in hardware, but it is obviously more expensive than \code{\_\_syncthreads()}. If such an operation existed, we could easily fold this into our API, thereby requiring no effort on behalf of users when upgrading to this new primitive.

\paragraph*{\textbf{Future Work}}

We see many avenues for future work, both for researchers and for hardware vendors. For researchers, using an auto-tuner to adjust the backoff variables for each specific GPU under differing loads of contention may prove worthwhile. Second, because semaphores have inherent dangers in their use, it is worthwhile to implement condition variables on the GPU\@. Third, we would like to extend our analysis to other GPU and GPU-like architectures, such as AMD GPUs and Intel MIC processors. In terms of future work from vendors, it would be useful to have a hardware global barrier (e.g. \code{\_\_syncblocks()}, mentioned above). This is something that a vendor must support explicitly, and there are many applications (e.g.\ FFT, Smith-Waterman, bitonic sort) that have shown a need for such a function~\cite{Xiao:2010:IBG}. And as much of the need for mutexes comes from managing shared queues, it would be very advantageous to have hardware-accelerated atomic queue intrinsics such as \code{enqueue()} and \code{dequeue()}.

%% file: appendix.tex
\vspace{1em}
\input{spinmutex_pseudocode}
\vspace{1em}
\input{famutex_pseudocode}
\vspace{1em}
\input{spinsem_pseudocode}
\vspace{1em}
\input{sleepsem_pseudocode}

%% file: spinmutex_pseudocode.tex
\begin{algorithm}

  \func GPU: SpinMutexLock(Mutex)
    \begin{algorithmic}[1]
      \STATE \textit{Acquired} $\leftarrow \FALSE$
      \WHILE{\textit{Locked} $ = \FALSE$}
        \STATE \textit{OldVal} $\leftarrow$ atomicExch(\textit{Mutex}, $1$)
        \IF{\textit{OldVal} $ = 0$}
          \STATE \textit{Acquired} $ \leftarrow \TRUE$
        \ELSIF{\textit{Acquired} $ = \FALSE \wedge $ \textit{UseBackoff} $ = \TRUE$}
          \STATE Backoff()
        \ENDIF
      \ENDWHILE
    \end{algorithmic}

  \vspace{5pt}

  \func GPU: SpinMutexUnlock(Mutex)
    \begin{algorithmic}[1]
      \STATE: \textit{Mutex} $ \leftarrow 0$
    \end{algorithmic}

  \caption{Spin-Lock Mutex Lock and Unlock functions. \label{listing:spinmutex}}

\end{algorithm}

%% file: famutex_pseudocode.tex
\begin{algorithm}

  \func GPU: FAMutexLock(Mutex)
    \begin{algorithmic}[1]
      \STATE \textit{TicketNumber} $\leftarrow$ atomicInc(\textit{Mutex.ticket})
      \WHILE{\textit{TicketNumber} $\neq$ \textit{Mutex.turn}}
        \STATE Backoff()
      \ENDWHILE
    \end{algorithmic}

  \vspace{5pt}

  \func GPU: FAMutexUnlock(Mutex)
    \begin{algorithmic}[1]
      \STATE: \textit{Mutex.turn} $\leftarrow$ \textit{Mutex.turn}$ + 1$
    \end{algorithmic}

  \caption{FA Mutex Lock and Unlock functions. \label{listing:famutex}}

\end{algorithm}

%% file: spinsem_pseudocode.tex
\begin{algorithm}

  \func GPU: SpinSemaphoreWait(Sem)
    \begin{algorithmic}[1]
      \STATE \textit{Acquired} $ \leftarrow \FALSE$
      \WHILE{\textit{Acquired} $ = \FALSE$}
        \STATE \textit{OldValue} $ \leftarrow $ atomicExch(\textit{Sem}, $0$)
        \IF{\textit{OldValue} $ > 1$}
          \STATE atomicExch(\textit{Sem}, \textit{OldValue} $ - 1$)
          \STATE \textit{Acquired} $ \leftarrow \TRUE$
        \ELSIF{\textit{OldValue} $ = 1$}
          \STATE atomicExch(\textit{Sem}, $1$)
        \ENDIF
        \IF{\textit{Acquired} $ = \FALSE \wedge $ \textit{UseBackoff} $ = \TRUE$}
          \STATE Backoff()
        \ENDIF
      \ENDWHILE
    \end{algorithmic}
  \vspace{5pt}

  \func GPU: SpinSemaphorePost(Sem)
    \begin{algorithmic}[1]
      \STATE \textit{Acquired} $ \leftarrow \FALSE$
      \WHILE{\textit{Acquired} $ = \FALSE$}
        \STATE \textit{OldValue} $ \leftarrow $ atomicExch(\textit{Sem}, $0$)
        \IF{\textit{OldValue} $ > 0$}
          \STATE atomicExch(\textit{Sem}, \textit{OldValue} $ + 1$)
          \STATE \textit{Acquired} $ \leftarrow \TRUE$
        \ENDIF
        \IF{\textit{Acquired} $ = \FALSE \wedge $ \textit{UseBackoff} $ = \TRUE$}
          \STATE Backoff()
        \ENDIF
      \ENDWHILE
    \end{algorithmic}
  \vspace{5pt}

  \caption{Spin-Lock Semaphore Wait and Post functions. \label{listing:spinsem}}

\end{algorithm}

%% file: sleepsem_pseudocode.tex
\begin{algorithm}

  \func GPU: SleepSemaphoreWait(Sem)
    \begin{algorithmic}[1]
      \STATE \textit{OldCount} $\leftarrow$ atomicInc(\textit{Sem.count})
      \IF{\textit{OldCount} $ < $ \textit{Sem.maxCount}}
        \STATE \textit{Acquired} $\leftarrow \TRUE$
      \ELSE
        \STATE \textit{Acquired} $\leftarrow \FALSE$
        \STATE \textit{WaitIndex} $\leftarrow$ atomicInc(\textit{Sem.ticket})
      \ENDIF
      \WHILE{$Acquired = \FALSE$}
        \IF{\textit{Sem.turn} $ > $ \textit{WaitIndex}}
          \STATE \textit{Acquired} $\leftarrow \TRUE$
        \ENDIF
      \ENDWHILE
    \end{algorithmic}
  \vspace{5pt}

  \func GPU: SleepSemaphorePost(Sem)
    \begin{algorithmic}[1]
      \STATE \textit{OldCount} $\leftarrow$ atomicDec(\textit{Sem.count})
      \IF{\textit{OldCount} $ > $ \textit{Sem.maxCount}}
        \STATE atomicInc(\textit{Sem.turn})
      \ENDIF
    \end{algorithmic}
  \vspace{5pt}

  \caption{Sleeping Semaphore Wait and Post functions. \label{listing:sleepsem}}

\end{algorithm}